\documentclass[reprint,showpacs,preprintnumbers,nofootinbib,amsmath,amssymb,aps,prd,floatfix,superscriptaddress]{revtex4-1}
\pdfoutput=1
\usepackage[utf8]{inputenc}
\usepackage{graphicx}
\usepackage{bm}
\usepackage{subfigure}
\usepackage{url}
\usepackage[hyperindex]{hyperref}
\usepackage{color}
\usepackage{xcolor}
\usepackage[ddmmyy,24hr]{datetime}
\usepackage{bigdelim}
\usepackage{booktabs}
\usepackage{dcolumn}
\usepackage{multirow}
\usepackage{subfigure}
\usepackage{cancel}
\usepackage{stackrel}
\usepackage{paralist}
\usepackage{xspace}
\usepackage{slashed}
\newcommand{\nua}[1]{\ensuremath{\rlap{\kern-2.5pt\ensuremath{\overset{\scriptscriptstyle(-)}{\phantom{\nu}}}}{\ensuremath{{\nu}_{#1}}}}}

\usepackage{enumitem}

\makeatletter
\def\namedlabel#1#2{\begingroup
    #2%
    \def\@currentlabel{#2}%
    \phantomsection\label{#1}\endgroup
}
\makeatother

\newcommand{\cenns}{CE$\nu$NS}
\newcommand{\cenas}{CE$\nu$AS}

\begin{document}

\title{Potentialities of a low-energy detector based on $^4$He evaporation to observe atomic effects in coherent neutrino scattering and physics perspectives}

\author{M. Cadeddu}
\email{matteo.cadeddu@ca.infn.it}
\affiliation{Università degli studi di Cagliari and Istituto Nazionale di Fisica Nucleare (INFN), Sezione di Cagliari,
Complesso Universitario di Monserrato - S.P. per Sestu Km 0.700,
09042 Monserrato (Cagliari), Italy}

\author{F. Dordei}
\email{francesca.dordei@cern.ch}
\affiliation{Istituto Nazionale di Fisica Nucleare (INFN), Sezione di Cagliari,
Complesso Universitario di Monserrato - S.P. per Sestu Km 0.700,
09042 Monserrato (Cagliari), Italy}

\author{C. Giunti}
\email{carlo.giunti@to.infn.it}
\affiliation{Istituto Nazionale di Fisica Nucleare (INFN), Sezione di Torino,
 Via P. Giuria 1, 
 I–10125 Torino, Italy}
 
\author{K. A. Kouzakov}
\email{kouzakov@gmail.com}
\affiliation{Department of Nuclear Physics and Quantum Theory of Collisions,
Faculty of Physics, Lomonosov Moscow State University, 
Moscow 119991, Russia}

\author{E. Picciau}
\email{emmanuele.picciau@ca.infn.it}
\affiliation{Università degli studi di Cagliari and Istituto Nazionale di Fisica Nucleare (INFN), Sezione di Cagliari,
Complesso Universitario di Monserrato - S.P. per Sestu Km 0.700,
09042 Monserrato (Cagliari), Italy}

\author{A. I. Studenikin}
\email{studenik@srd.sinp.msu.ru}
\affiliation{Department of Theoretical Physics, Faculty of
Physics, Lomonosov Moscow State University, Moscow 119991, Russia}
\affiliation{Joint Institute for Nuclear Research, Dubna 141980, Moscow Region, Russia}

\begin{abstract}
We propose an experimental setup to observe coherent elastic neutrino-atom scattering (\cenas) using electron antineutrinos from tritium decay and a liquid helium target. In this scattering process with the whole atom, that has not beeen observed so far, the electrons tend to screen the weak charge of the nucleus as seen by the electron antineutrino probe. The interference between the nucleus and the electron cloud produces a sharp dip in the recoil spectrum at atomic recoil energies of about 9~meV, reducing sizeably the number of expected events with respect to the coherent elastic neutrino-nucleus scattering case. We estimate that with a 60~g tritium source surrounded by 500~kg of liquid helium in a cylindrical tank, one could observe the existence of \cenas\ processes at 3$\sigma$ in 5 years of data taking. Keeping the same amount of helium and the same data-taking period, we test the sensitivity to the Weinberg angle and a possible neutrino magnetic moment for three different scenarios: 60~g, 160~g, and 500~g of tritium. In the latter scenario, the Standard Model (SM) value of the Weinberg angle can be measured with a statistical uncertainty of $\sin^2{\vartheta_W^{\mathrm{SM}}}^{+0.015}_{-0.016}$. This would represent the lowest-energy measurement of $\sin^2{\vartheta_W}$, with the advantage of being not affected by the uncertainties on the neutron form factor of the nucleus as the current lowest-energy determination. Finally, we study the sensitivity of this apparatus to a possible electron neutrino magnetic moment and we find that using 60~g of tritium it is possible to set an upper limit of about $7\times10^{-13}\,\mu_B$ at 90\% C.L., that is more than one order of magnitude smaller than the current experimental limit.
\end{abstract}


\maketitle

\section{INTRODUCTION}

Coherent elastic neutrino-nucleus scattering (\cenns), has been recently observed by the COHERENT experiment~\cite{Akimov:2017ade,Akimov:2018vzs}, after many decades from its prediction~\cite{PhysRevD.9.1389,Freedman:1977xn,Drukier:1983gj}. This observation triggered a lot of attention from the scientific community and unlocked a new and powerful tool to study many and diverse physical phenomena: nuclear physics~\cite{Cadeddu:2017etk,Papoulias:2019lfi}, neutrino properties~\cite{Coloma:2017ncl,Miranda:2019wdy,Cadeddu:2018dux}, physics beyond the Standard Model (SM)~\cite{Dutta:2019eml,Miranda:2019skf,Heeck:2018nzc,Abdullah:2018ykz,Farzan:2018gtr, Liao:2017uzy,Kosmas:2017tsq}, and electroweak (EW) interactions~\cite{Cadeddu:2018izq,Canas:2018rng}. The experimental challenge related to the \cenns\ observation is due to the fact that in order to meet the coherence requirement $q R \ll 1$~\cite{Bednyakov:2018mjd},
where $q=|\vec{q}|$ is the three-momentum transfer and $R$ is the nuclear radius, one has to detect very small nuclear recoil energies $E_R$, lower  than  a  few keV.

At even lower momentum transfers, such that $q R_{\mathrm{atom}} \ll 1$, where $R_{\mathrm{atom}}$ is the radius of the target atom including the electron shells, the reaction can be viewed as taking place on the atom as a whole. In Ref.~\cite{Sehgal1986} it has been noted that, in the case of electron neutrino scattering, the interference between the electron cloud and the nucleus becomes destructive. As a consequence, the scattering amplitude becomes null and changes sign as $q^2$ varies from large to small values, producing a sharp dip in the differential cross section. In practice, electrons tend to screen the weak charge of the nucleus as seen by an electron neutrino probe. This effect should be visible for momentum transfers of the order of $q^2\sim100\,\mathrm{keV}^2$, corresponding to atomic recoil energies $T_R \sim 10\,\mathrm{meV}$. For recoil energies of  $\mathcal{O}(100\,\mathrm{meV})$ this effect starts to become completely negligible. The small recoils needed are well below the thresholds of detectability  of currently available detectors, making it very difficult to observe this effect. 

However, in Ref.~\cite{Maris:2017xvi} a new technology based on the evaporation of helium atoms from a cold surface and their subsequent detection using field ionization has been proposed for the detection of low-mass dark matter particles. In this configuration, the nuclear recoils induced by dark-matter scattering produce elementary excitations (phonons and rotons) in the target that can result in the evaporation of helium atoms, if the recoil energy is greater than the binding energy of helium to the surface. Given that the latter can be below 1~meV, this proposed technique represents an ideal experimental setup to observe atomic effects in coherent neutrino scattering. 

Here, we propose a future experiment that would allow the observation of coherent elastic neutrino-atom scattering (\cenas) processes and we investigate its sensitivity. Since this effect could be visible only for extremely small recoil energies, in order to achieve a sufficient number of low-energy \cenas\ events, electron neutrinos with energies of the order of a few
keV need to be exploited. Unfortunately, there are not so many available sources of such low-energy neutrinos. In this paper, we investigate the possibility to use a tritium $\beta$-decay source, that is characterized by a Q-value of 18.58~keV. The PTOLEMY project~\cite{Betts:2013uya,Betti:2019ouf}, that aims to develop a scalable design to detect cosmic neutrino background, is already planning to use about 100~g of tritium, so we can assume that a similar or even larger amount could be available in the near future. Moreover, we show the potentialities of such a detector to perform the lowest-energy measurement of the weak mixing angle $\vartheta_W$, also known as the Weinberg angle, a fundamental parameter in the theory of Standard Model electroweak interactions, and to reveal a magnetic moment of the electron neutrino
below the current limit. 

\section{ATOMIC EFFECTS IN COHERENT SCATTERING}

In order to derive the cross section for a \cenas\ process $\nu_\ell + A \to \nu_\ell +A$, where A is an atom and $\ell = e, \mu, \tau$, we start from the differential-scattering cross-section of a \cenns\ process as a function of the recoil energy $E_R$ of the nucleus~\cite{Sehgal1986,gaponov}
\begin{equation}
    \frac{d\sigma^{\mathrm{CE}\nu\mathrm{NS}}}{dE_{R}}=\frac{G_F^2}{\pi} C_{V}^2 m_N \left(1-\frac{m_N E_R}{2 E_{\nu}^2} \right) \,,
    \label{eq:sigmadiff}
\end{equation}
 where $G_F$ is the Fermi constant, $m_N$ is the nuclear mass, $E_{\nu}$ is the neutrino energy and $C_V$ is the $q^2$-dependent matrix element of the vector neutral-current charge
 \begin{equation}
    C_V =\frac{1}{2}[(1-4\sin^2\vartheta_W)Z\,F_{\mathrm{Z}}(q^2) - N\,F_{\mathrm{N}}(q^2)]\,.
    \label{eq:cvnucleus}
\end{equation}
Here, $Z$ and $N$ are the number of protons and neutrons in the atom and $F_{\mathrm{N}}(q^2)$ ($F_{\mathrm{Z}}(q^2)$) is the nuclear neutron (proton) form factor~\cite{PhysRevC.86.024612,Barranco:2005yy,Cadeddu:2017etk}.  In principle, one should also consider the axial coupling $C_A$ contribution, but for even-even (spin zero) nuclei it is equal to zero. Since in this paper we take as a target a $^4$He detector, we have $C_A=0$.\\

As anticipated in the introduction, when the energy of the incoming neutrino is low enough, atomic effects arise. 
In the case of \cenas\ processes, $d\sigma^{\mathrm{CE}\nu\mathrm{AS}}/dT_R$ can be derived starting from the formula in Eq.~(\ref{eq:sigmadiff}) with the inclusion of the electron contribution to the vector coupling~\cite{Sehgal1986,gaponov}
\begin{equation}
    C_V^{^{\mathrm{Atom}}} = C_V + \frac{1}{2}(\pm 1 + 4\sin^2\vartheta_W)Z\,F_{\mathrm{e}}(q^2)\,,
\label{eq:cvatom}
\end{equation}
where $F_{\mathrm{e}}(q^2)$ is the electron form factor and the $+$ sign applies to ${\nu}_e$ and $\bar{\nu}_e$, while the $-$ sign applies to all the other neutrino species. As explained in Ref.~\cite{Sehgal1986}, the $+$ sign is responsible for the destructive interference between the electron and nuclear contributions. In principle, one should add also the axial contribution from the electron cloud, $C_A^{^{\mathrm{Atom}}}$, that however becomes null when the number of spin up and down electrons is the same, which applies to our scenario. Moreover, at the low momentum transfers considered in \cenas\ processes, one can safely put $F_{\mathrm{N}}(q^2)=F_{\mathrm{Z}}(q^2) = 1$. This allows to derive physics properties from the analysis on \cenas\ processes being independent on the knowledge of the neutron distribution that is largely unknown~\cite{Cadeddu:2017etk,AristizabalSierra:2019zmy}. 

\begin{figure}[h!]
\centering
\includegraphics[scale=0.45]{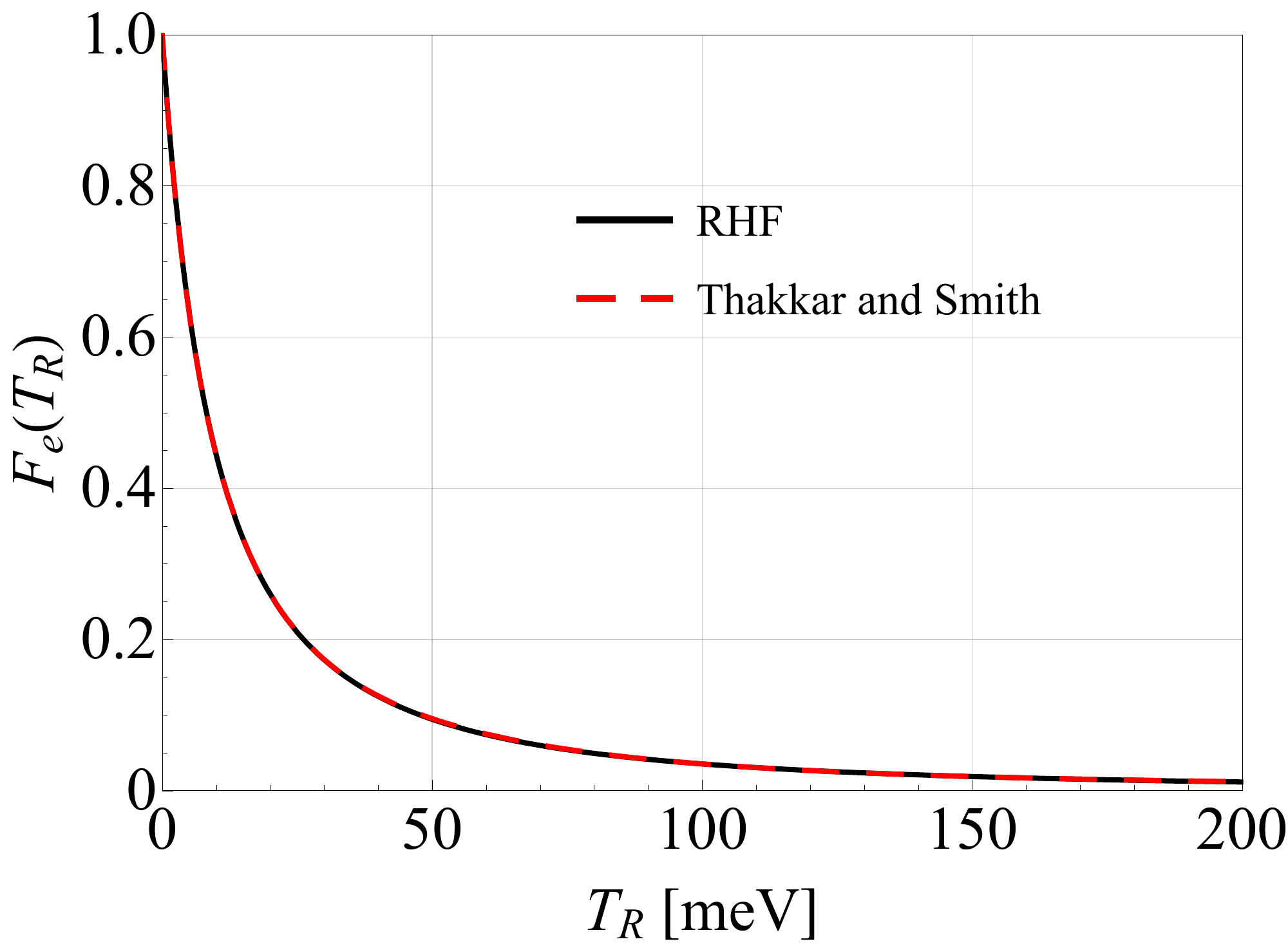}
\caption{The He electron form factor as a function of the recoil energy $T_R$. The solid (black) curve represents the RHF model~\cite{Coulthard1967} and the dashed (red) curve the strongly correlated model of Thakkar and Smith~\cite{Thakkar1977}. \label{fig:He_scattering_factor}}
\end{figure}

As visible from Eq.~(\ref{eq:cvatom}), in \cenas\ processes a key role is played by the electron form factor, that is defined as the Fourier transform of the electron density of an atom. In contrast to the case of atomic hydrogen, the He electron density is not known exactly. In our present study we employ the following parameterisation of the He electron form factor, that has proved to be particularly effective and accurate (see, for instance, Refs.~\cite{Doyle1968,Brown,Thakkar1992})
\begin{equation}
    F_{\mathrm{e}}(q^2)= \mathcal{A} \cdot \left( \, \sum_{i=1}^{4} a_i\cdot e^{-b_i\left(q/4\pi\right)^2} + c \right)\,.
\end{equation}
The parameters $a_i=\{0.8734,0.6309,0.3112,0.178\}$, $b_i=\{9.1037,3.3568,22.9276,0.9821\}$ and $c=0.0064$, extracted from Tab. 6.1.1.4 of Ref.~\cite{Brown}, are given by a close fit of the numerical calculations for $F_{\mathrm{e}}(q^2)$ using a well-established theoretical model of the He electron wave function. The normalization $\mathcal{A}$ is a scaling factor such that \mbox{$F_{\mathrm{e}}(q^2)\rightarrow1$} for $q\to0$, as in the nuclear form factor definition. This parameterisation assumes that the electron density is spherically symmetric so that the value of the Fourier transform only depends on the distance from the origin in reciprocal space. The moment transfer $q$ is related to the atomic recoil energy through $q=\sqrt{2 m_A T_R}$, where $m_A\sim m_N$ is the atomic mass.\footnote{Here we neglect the binding of the atom in the liquid helium target, since it becomes of relevance only at the energy scale $T_R\lesssim1$~meV.} For various models of the He electron wave function that give an accurate value of the electron binding energy, the numerical results for $F_{\mathrm{e}}(q^2)$ are known to be practically identical in a wide range of $q$ values. To illustrate this feature, we compare in Fig.~\ref{fig:He_scattering_factor} two He electron form factors obtained with two different well-known approaches: (i) the Roothaan-Hartree-Fock (RHF) method~\cite{Coulthard1967} and (ii) the variational method using a strongly correlated ansatz~\cite{Thakkar1977}. Despite the fact that the indicated methods treat electron-electron correlations in an opposite manner (the first completely neglects them, while the second takes them into account explicitly), they yield very close electron form-factor values. This illustration shows a negligible role of the theoretical uncertainties associated with the He electron form factor in the analysis carried out in the next sections.  

\section{EXPERIMENTAL SETUP AND EXPECTED NUMBER OF EVENTS}
\label{sec:detector}

As mentioned in the introduction, a tritium source would provide electron antineutrinos in the needed energy range. Indeed, tritium $\beta^-$ decay produces one electron, one antineutrino and a $^3\mathrm{He}$ atom via the decay $^3\mathrm{H} \rightarrow \ ^3\mathrm{He} + e^- + \bar{\nu}_e$, with a lifetime $\tau=17.74$ years. The antineutrino energy spectrum ranges from 0~keV to the Q-value, with a maximum at approximately 15~keV~\cite{Giunti:2007ry}. 
The number of neutrinos released, $N_\nu$, after a time $t$ follows the simple exponential decay law 
\begin{equation}
N_{\nu}(t) = N_{_{^3\mathrm{H}}} \left(1-e^{-t/\tau}\right),
\end{equation}
where $N_{_{^3\mathrm{H}}}$ is the number of tritium atoms in the source at $t=0$. 
Given the rather large tritium lifetime, the antineutrino rate is expected to stay almost constant in the first 5 years.   

\begin{figure}[h!]
\centering
\includegraphics[scale=0.27]{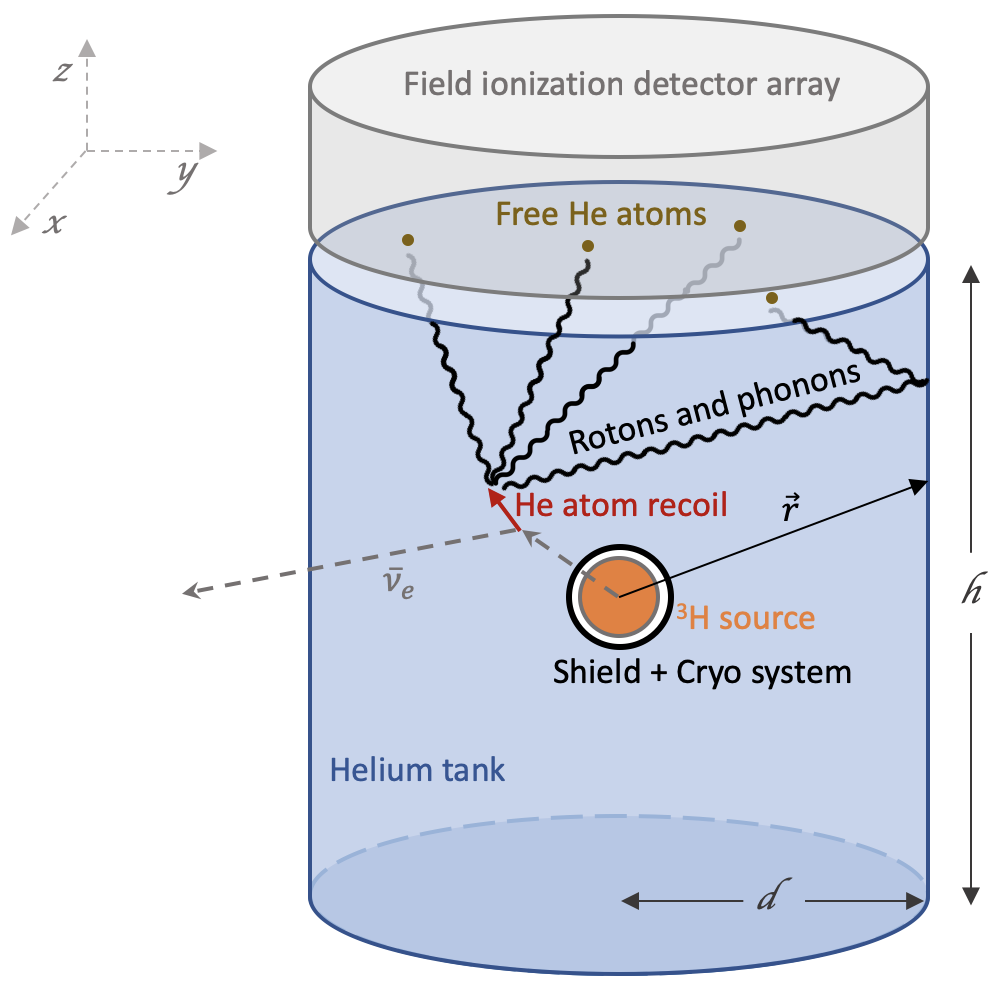}
\caption{Schematic representation of the detector proposed to observe \cenas\ processes. The recoil of a helium atom after the scattering with an electron antineutrino coming from the tritium source in the center produces phonons and rotons which, upon arrival at the top surface, cause helium atoms to be released by quantum evaporation. A field ionization detector array on the top surface, as proposed in Ref.~\cite{Maris:2017xvi}, detects the number of helium atoms evaporated. \label{fig:cyl_conf}}
\end{figure}

We consider a detector setup such that the tritium source is surrounded with a cylindrical superfluid-helium tank, as depicted in Fig.~\ref{fig:cyl_conf}. This configuration allows to maximize the geometrical acceptance, while allowing to have a top flat surface where helium atoms could evaporate after a recoil. This surface should be equipped as described in Ref.~\cite{Maris:2017xvi}, to detect the small energy deposited by the helium evaporation. In order to shield the helium detector from the electrons produced by the tritium decay, an intermediate thin layer of heavy material has to be inserted between the source and the detector, which also acts as a vessel for the source. Cautions have to be taken such that all the materials used in the detector are extremely radiopure, such that the background contaminations are kept under control. Finally, the energy deposited by the electrons in the shield would cause a significant heating of the surrounding helium. Thus the shield has to be placed inside a cryocooler to keep the source at the desired temperature, surrounded by a vacuum layer to further isolate the source.

The expected \cenas\ differential rate $d\mathcal{N}/(dt\,dT_R)$ in such a cylindrical configuration, that represents the number of neutrino-induced events $\mathcal{N}$ that would be observed in the detector each second and per unit of the atomic recoil energy, is 
\begin{equation}
    \frac{d\mathcal{N}}{dt\,dT_R}=n \int_{V} \int_{E_\nu^{\mathrm{min}}}^Q \frac{1}{4\pi r^2}\ \frac{dN_\nu}{dt\ dE_\nu} \frac{d\sigma^{\mathrm{CE}\nu\mathrm{AS}}}{dT_R}\  dV\ dE_\nu \,,
\label{eq:diff_rate}
\end{equation}
where $n$ is the number density of helium atoms in the target, $dV$ is the infinitesimal volume around the position $\overrightarrow{r} \equiv (x,y,z)$ in the detector, $dN_\nu/(dt\ dE_\nu)$ is the differential neutrino rate and \mbox{$E_\nu^{^{\mathrm{min}}} \simeq \sqrt{m_A T_R/2}$} is the minimum antineutrino energy necessary to produce an atom recoil of energy $T_R$. Note that, if the atomic effect is neglected, the differential number of events $d\mathcal{N}^{\mathrm{CE}\nu\mathrm{NS}}/dE_R$ could be  straightforwardly obtained with the following set of substitutions:  $d\sigma^{\mathrm{CE}\nu\mathrm{AS}}/dT_R \to d\sigma^{\mathrm{CE}\nu\mathrm{NS}}/dE_R$, $m_A \to m_N$ and $T_R \to E_R$. However, given that $m_A \simeq m_N$, the atomic and nuclear recoil energies are practically coincident, $T_R \simeq E_R$. 

To illustrate the consequences of the atomic effect on the expected number of events, we consider a detector of height $h=160$ cm, and radius $d=90$ cm filled with \mbox{500 kg} of helium, a tritium source of 60 g, and a data-taking period of 5 years. The choice of this particular configuration will be clarified in Sec.~\ref{sec:sensiticity_atomic}. 

\begin{figure}[h!]
\centering
\includegraphics[scale=0.33]{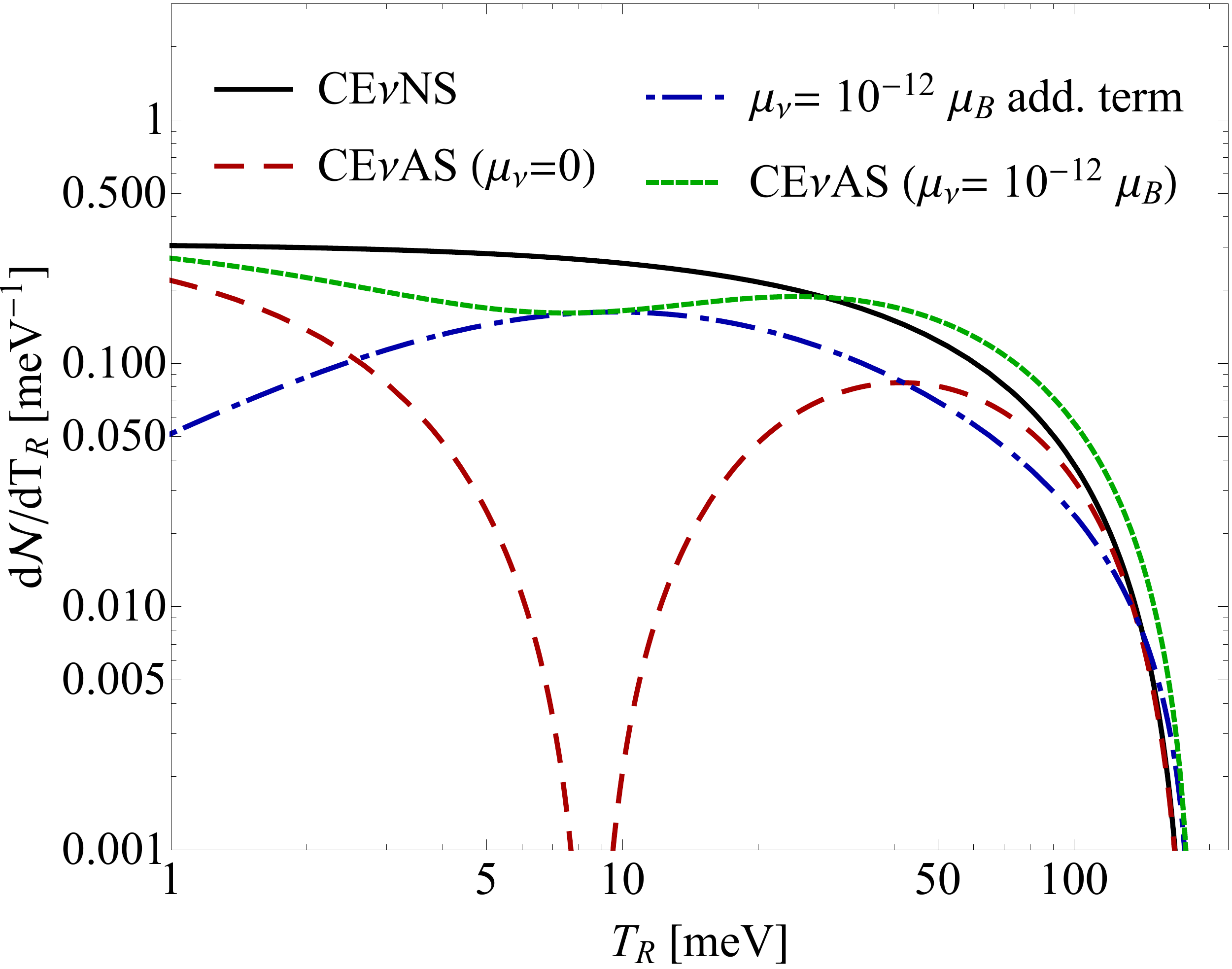}
\caption{Differential number of neutrino-induced events as a function of the atomic recoil energy $T_R$ in a logarithmic scale on both axes. The \cenns\ differential number is shown by the black solid line while the \cenas\ one is shown by the dashed red line. The dashed-dotted blue line represents the additional term appearing in the \cenas\ differential number of events assuming a neutrino magnetic moment of $\mu_\nu = 10^{-12}\,\mu_B$  while the dotted green line represents the total differential number of \cenas\ for the same value of $\mu_\nu$ (i.e. using the differential cross section in Eq.~\ref{magneticmomentcontr}) .  \label{fig:diff_rate}}
\end{figure}

The number of neutrino-induced events as a function of the recoil energy $T_R$ is obtained by integrating 
the differential rate defined in Eq.~(\ref{eq:diff_rate}) for a time period of 5 years. The result is illustrated in Fig.~\ref{fig:diff_rate}, where the \cenns\ differential rate is shown by the black solid line, while the \cenas\ one is shown by the dashed red line. As stated in the introduction, when atomic effects are considered, electrons screen the weak charge of the nucleus as seen by the antineutrino. The screening is complete for atomic recoil energies $T_R$ such that $C_V^{^{\mathrm{Atom}}}=0$, or, in accordance with Eqs.~(\ref{eq:cvnucleus}) and~(\ref{eq:cvatom}), when
\begin{equation}
    F_e(T_R)=\frac{\frac{N}{Z}-(1-4 \sin^2\vartheta_W)}{1+4\sin^2\vartheta_W}\,,
    \label{eq:screening_condition}
\end{equation}
where for $^{4}\mathrm{He}$ atoms $N/Z=1$. Using the SM prediction of the Weinberg angle at near zero momentum transfer $\sin^2{\vartheta_W^{\mathrm{SM}}} = 0.23857(5)$~\cite{Tanabashi:2018oca}, calculated in the $\overline{\mathrm{MS}}$ scheme, we obtain from Eq.~(\ref{eq:screening_condition}) the following condition: $F_e(T_R)=0.4883$. Thus, the screening is complete for \mbox{$T_R \simeq 9$ meV} (see Fig.~\ref{fig:He_scattering_factor}). Due to this destructive interference between the nuclear and the electron contributions, the number of events drops rapidly to zero, as shown by the red dashed line in Fig.~\ref{fig:diff_rate}.

\section{SENSITIVITY TO THE ATOMIC EFFECT}
\label{sec:sensiticity_atomic}

We test the feasibility to observe the atomic effect in coherent neutrino scattering using the experimental setup described in Sec.~\ref{sec:detector}. For this purpose, we build the following least-squares function
\begin{equation}
    \chi^2=\left( \frac{\mathcal{N}^{\mathrm{CE}\nu\mathrm{AS}}-\mathcal{N}^{\mathrm{CE}\nu\mathrm{NS}}}{\sigma}\right)^2\,,
\label{chi1}
\end{equation}
where $\mathcal{N}^{\mathrm{CE}\nu\mathrm{AS}}$ represents the number of neutrino-induced events observed considering the atomic effect, while $\mathcal{N}^{\mathrm{CE}\nu\mathrm{NS}}$ is the number of events expected ignoring such an effect, which represents our reference model~\cite{Cowan:2010js}. These quantities are obtained integrating in time and recoil energy the differential spectrum in Eq.~(\ref{eq:diff_rate}), considering for the latter the range of 1-184 meV, being the upper limit the maximum recoil energy that a neutrino with energy $E_{\nu}$ can give to an atom with mass $m_A$, which is given by
\begin{equation}
    T_R^{\mathrm{max}}=\frac{2  E_{\nu}^2}{m_A+2E_{\nu}}\,.
\end{equation}
Assuming that the main uncertainty contribution is due to the available statistics,
the denominator in Eq.~(\ref{chi1}) is set to be $\sigma = \sqrt{\mathcal{N}^{\mathrm{CE}\nu\mathrm{AS}}}$. 

In order to find the experimental setup that would allow to reach a sensitivity of at least 3$\sigma$, we calculate this least-squares function in terms of three parameters: the amount of helium in the tank, the amount of tritium in the source, and the time of data-taking (the main contributor being the amount of tritium used). We find that a reasonable combination of these parameters that would allow the observation of the atomic effect is
\mbox{500 kg} of helium, 60 g of tritium, and 5 years of data-taking. In this scenario, the expected number of \cenas\ events is $\mathcal{N}^{\mathrm{CE}\nu\mathrm{AS}}=6.7$, to be compared with the
expected number of \cenns\ events $\mathcal{N}^{\mathrm{CE}\nu\mathrm{NS}}=14.6$.

\begin{figure}[th!]
\centering
\includegraphics[scale=0.33]{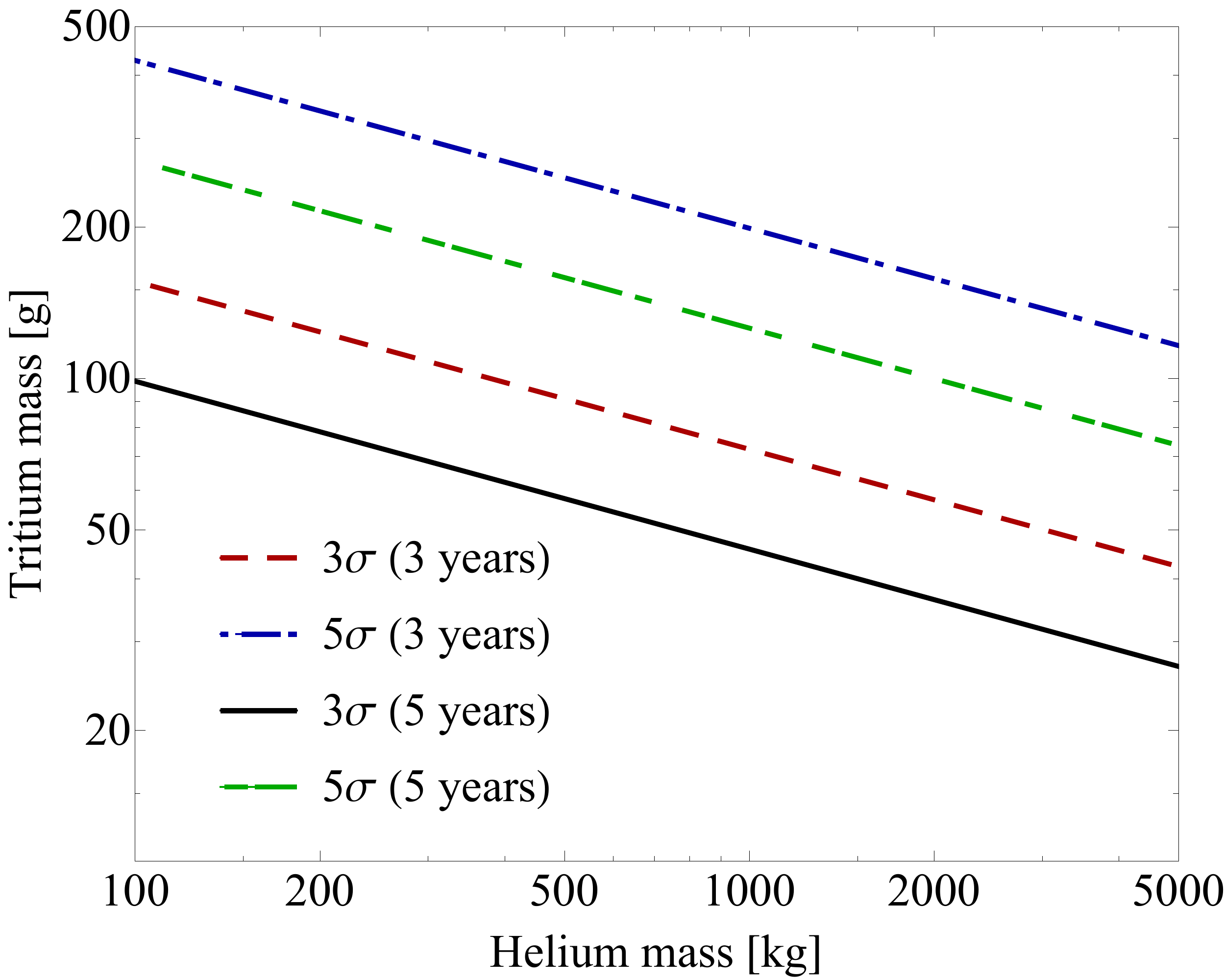}
\caption{Iso-sigma curves to observe (3$\sigma$ in dashed red and solid black) or discover (5$\sigma$ in dotted-dashed blue and long dashed green) the atomic effect, as a function of the helium and tritium masses, considering a data taking time of 3 and 5 years, respectively. \label{fig:sensatomic}}
\end{figure}

In order to claim a discovery, i.e. reach a sensitivity of 5$\sigma$, the amount of tritium needed increases to 160 g, leaving the other parameters unchanged. The expected number of events in this case becomes $\mathcal{N}^{\mathrm{CE}\nu\mathrm{AS}}=17.7$
considering the atomic effect and $\mathcal{N}^{\mathrm{CE}\nu\mathrm{NS}}=38.9$
without the atomic effect. One can note that the atomic screening reduces the number of events by almost one half with the particular experimental setup proposed in this paper.
For completeness, in Fig.~\ref{fig:sensatomic} one can see the amount of tritium and helium mass needed to observe (3$\sigma$) or discover (5$\sigma$) the atomic effect, considering a data taking time of 3 and 5 years. 

Clearly, all these estimates are to be refined for a specific experiment by taking into account systematic contributions from backgrounds and detector efficiencies and resolutions. However, neither of the indicated contributions appear to be seriously limiting~\cite{Hertel:2018arXiv} and, hence, the conclusions drawn in this paper should remain valid.
\\

\section{PHYSICS PERSPECTIVES}

In the following, we assume that \cenas\ has been observed and we
discuss the sensitivity of the determination of the Weinberg angle and the neutrino magnetic moment.
Motivated by the studies performed in the previous section,
we consider a detector made of 500 kg of helium, 5 years of data-taking, and three different scenarios for the source: 60 g, 160 g, and 500 g of tritium. The last scenario is considered in order to see the potentialities of such a detector if a large quantity of tritium will become available. For completeness, the expected number of events in this optimistic scenario becomes $\mathcal{N}^{\mathrm{CE}\nu\mathrm{AS}}=55$
taking into account the atomic effect
and $\mathcal{N}^{\mathrm{CE}\nu\mathrm{NS}}=122$ without it.

\subsection{DETERMINATION OF THE WEINBERG ANGLE}

Since the vector coupling $C_V^{^{\mathrm{Atom}}}$ in Eq.~(\ref{eq:cvatom})
depends on $\sin^2\vartheta_W$,
the Weinberg angle can be measured in $\mathrm{CE}\nu\mathrm{AS}$
processes.
In order to quantify the sensitivity of a measurement
of the Weinberg angle with the experimental setup described above,
we consider a deviation of $\sin^2\vartheta_W$ from the
Standard Model value $\sin^2{\vartheta_W^{\mathrm{SM}}}$
in the least-squares function
\begin{equation}
\label{eq:chi2sintheta}
    \chi^2(\sin^2\vartheta_W)=\left(\frac{\mathcal{N}^{\mathrm{CE}\nu\mathrm{AS}}_{\mathrm{SM}}-\mathcal{N}^{\mathrm{CE}\nu\mathrm{AS}}(\sin^2\vartheta_W)}{\sqrt{\mathcal{N}^{\mathrm{CE}\nu\mathrm{AS}}_{\mathrm{SM}}}}\right)^2.
\end{equation}
Here, $\mathcal{N}^{\mathrm{CE}\nu\mathrm{AS}}_{\mathrm{SM}}$ represents the expected number of \cenas\ events if
$\sin^2\vartheta_W=\sin^2{\vartheta_W^{\mathrm{SM}}}$, while $\mathcal{N}^{\mathrm{CE}\nu\mathrm{AS}}(\sin^2\vartheta_W)$ is the number of events for a given value of the Weinberg angle. \\

\begin{figure}[th!]
\centering
\includegraphics[scale=0.30]{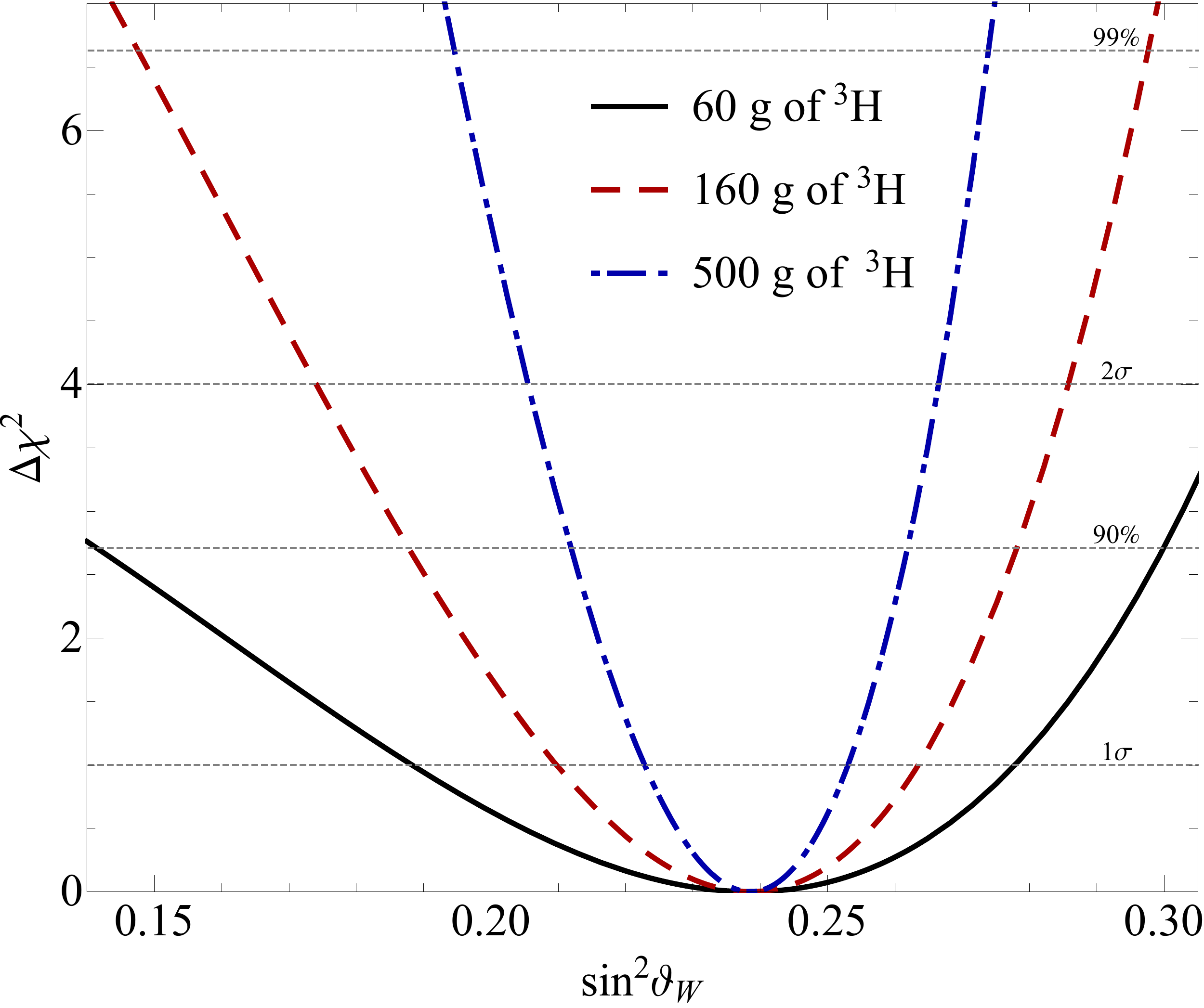}
\caption{With  the  black solid  curve  it  is  shown  the  \mbox{$\Delta \chi^2 = \chi^2 - \chi^2_{\mathrm{min}}$}, where the $\chi^2$ is defined in Eq.~(\ref{eq:chi2sintheta}), as a function of the Weinberg angle $\sin^2{\vartheta_W}$, obtained considering a tritium source of 60 g, while with the dashed red line and the dotted-dashed blue line it is shown the $\Delta \chi^2$ obtained considering a tritium source of 160 g and 500 g, respectively. \label{fig:sin2theta}}
\end{figure}

Figure~\ref{fig:sin2theta} shows the $\Delta \chi^2 = \chi^2 - \chi^2_{\mathrm{min}}$ profile as a function of $\sin^2{\vartheta_W}$ for the three different scenarios described above. The black solid, red dashed, and blue dotted-dashed lines refer to 60 g, 160 g, and 500 g of tritium, respectively.
The uncertainties achievable in the three scenarios are $^{+0.04}_{-0.05}$, $^{+0.025}_{-0.029}$ and $^{+0.015}_{-0.016}$, respectively.

\begin{figure}[t!]
\centering
\includegraphics[scale=0.24]{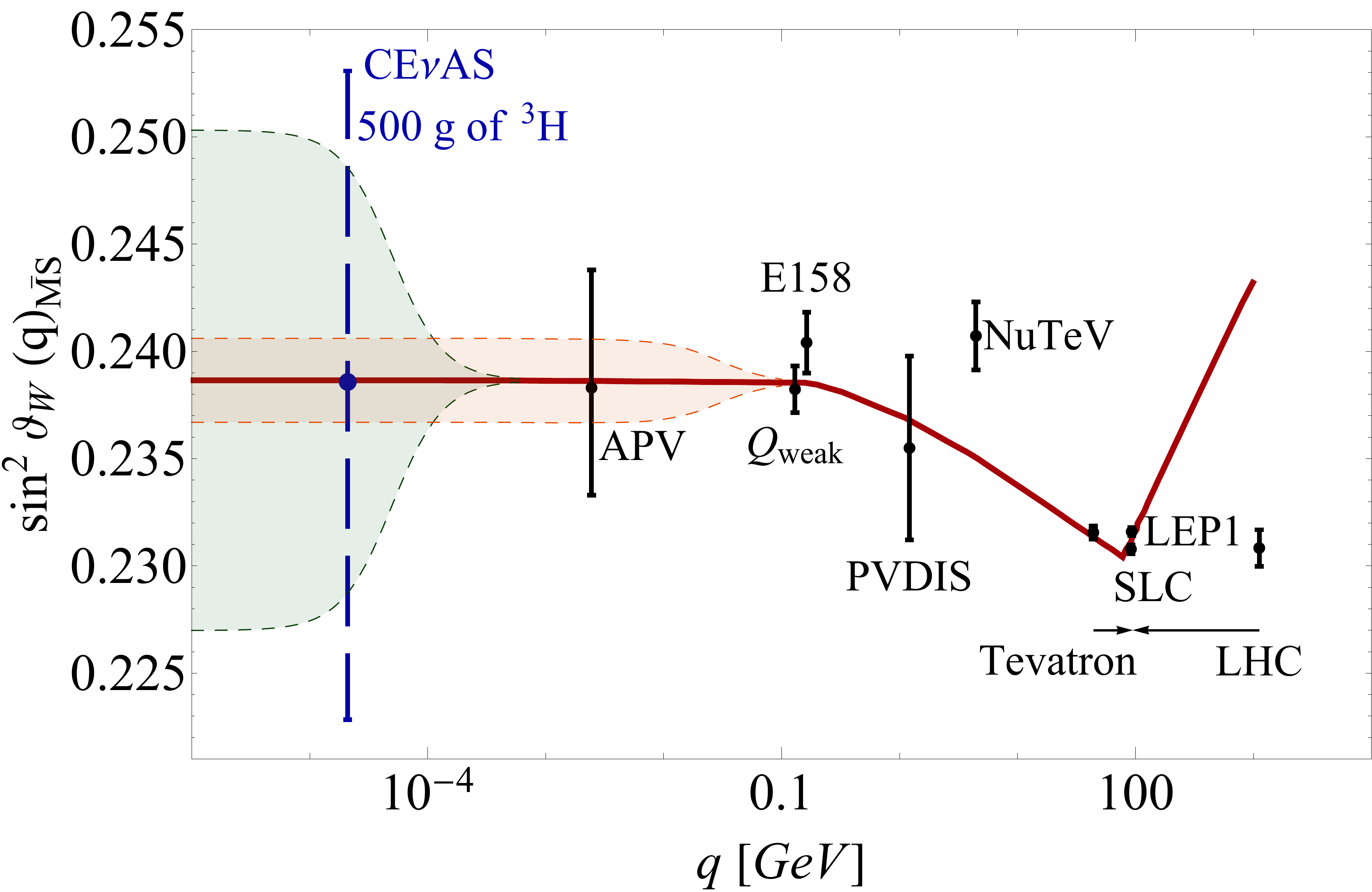}
\caption{Variation of $\sin^2 \vartheta_{\text{W}}$ with energy scale $q$. The SM prediction is shown as the red solid curve, together with
experimental determinations in black at the $Z$-pole~\cite{Tanabashi:2018oca} (Tevatron, LEP1, SLC, LHC),
from atomic parity violation on caesium~\cite{Wood:1997zq,Dzuba:2012kx,Cadeddu:2018izq}, which has a typical momentum transfer given by $\langle q\rangle\simeq$~2.4 MeV, M{\o}ller scattering~\cite{Anthony:2005pm} (E158), deep inelastic scattering of polarized electrons on deuterons~\cite{Wang:2014bbc} ($ e^2H $ PVDIS) and from
neutrino-nucleus scattering~\cite{Zeller:2001hh} (NuTeV) and the result from the proton's weak charge 
at $q = 0.158$ GeV~\cite{Androic:2018kni} ($ Q_{weak} $). For clarity the  Tevatron  and  LHC points have been displayed horizontally to the left and to the right, respectively, as indicated by the arrows. In dashed-blue it is shown the result that could be achieved using the experimental setup proposed in this paper, obtained exploiting \cenas\ effect at very low momentum transfer. The orange and green regions indicate the values of the weak mixing angle that are obtained for particular masses and couplings of a hypothetical $Z_d$ boson, see the text for more details. \label{fig:sin2all}}
\end{figure}

Figure~\ref{fig:sin2all} shows a summary of the  weak  mixing  angle  measurements  at different values of $q$, together with the SM prediction. The uncertainty that can be reached with the method proposed in this paper
in the case of a 500~g tritium source
is shown by the dashed blue line.
Despite the fact that this uncertainty is rather large, such measurement would represent a unique opportunity to explore the low-energy sector, since the value $\langle q\rangle\simeq 2 \times 10^{-5}$~GeV is several orders of magnitude smaller than in all the other measurements. Given that the value of the Weinberg angle provides a direct probe of physics phenomena not included in the SM, such a measurement would give complementary information to those at mid and high-energy. In particular, this measurement would be highly  sensitive  to an extra dark boson, $Z_d$, whose existence is predicted  by  grand  unified  theories, technicolor models,  supersymmetry  and  string  theories~\cite{Safronova:2017xyt}.
Measurements of the Weinberg angle provide constraints on the properties of this dark boson as its mass, $m_{Z_d}$, its kinetic coupling to the SM fermions, $\varepsilon$, and its $Z$-$Z_d$ mass-mixing coupling, $\delta$~\cite{Davoudiasl:2012ag,Davoudiasl:2012qa,Davoudiasl:2015bua}.
As an example, the orange and green regions in Fig.~\ref{fig:sin2all} show the low-$q$ deviations of
$\sin^2 \vartheta_{\text{W}}$
with respect to the SM value
predicted for two particular configurations of these parameters:
the contours of the orange shadowed region correspond to
$m_{Z_d}=30$~MeV, $\delta=0.015$, and $\varepsilon=1\times10^{-3}$,
while the contours of the green shadowed region have been obtained using $m_{Z_d}=0.05$~MeV, $\delta=0.0015$, and $\varepsilon=1\times10^{-5}$.
These values have been chosen such that $|\varepsilon\, \delta| \leq 8\times 10^{-4}$~\cite{Davoudiasl:2015bua}, which allows to roughly satisfy the existing upper bounds on these quantities\footnote{Note that in Ref.~\cite{Davoudiasl:2015bua} the constraint is expressed in terms of $\delta'$, but for our choice of parameters $\delta'\simeq\delta$.}. The shadowed regions indicate the values of the weak mixing angle obtained leaving the mass of the $Z_d$ boson invariant, but using smaller values of $\varepsilon$ and $\delta$. As it is visible from Fig.~\ref{fig:sin2all}, the impact of the hypothetical $Z_d$ boson starts at values of the transferred momentum equal to its mass and extends to lower values. Thus, a measurement like the one proposed in this paper would be useful to better constraint even lighter $Z_d$'s. 

\subsection{EFFECT OF NEUTRINO MAGNETIC MOMENT}

The experiment proposed in this paper would be highly sensitive to a possible neutrino magnetic moment.
So far, the most stringent constraints on the electron neutrino magnetic moment in laboratory experiments have been obtained looking for possible  distortions  of  the  recoil electron energy spectrum in neutrino-electron elastic scattering, exploiting solar neutrinos and reactor anti-neutrinos. The Borexino collaboration reported the best current limit on the effective magnetic moment of $2.8\times 10^{-11} \mu_B$ at 90\% confidence level (C.L.) using the electron recoil spectrum from $^{7}\mathrm{Be}$ solar neutrinos~\cite{Borexino:2017fbd}. The best magnetic moment limit from reactor anti-neutrinos, obtained by the GEMMA experiment, is very similar and it corresponds to $2.9\times 10^{-11} \mu_B$ (90\% C.L.)~\cite{Beda:2013mta}.\\

In the original SM with massless neutrinos the neutrino magnetic moments are vanishing,
but the results of neutrino oscillation experiments
have proved that the SM must be extended in order to give masses to the neutrinos. In the minimal extension of the SM in which neutrinos acquire Dirac masses through the introduction of right-handed neutrinos, the neutrino magnetic moment is given by~\cite{PhysRevLett.45.963,PhysRevD.24.1883,Kayser:1982br,PhysRevD.26.3152,PhysRevD.25.766,Shrock:1982sc}
\begin{equation}
    \mu_{\nu} = \frac{3\, e \,G_F}{8 \sqrt{2}\, \pi^2 } m_\nu \simeq 3.2\times10^{-19}\left( \frac{m_\nu}{ \mathrm{eV}} \right) \mu_B\,,
\label{ESMmunu}
\end{equation}
where $\mu_B$ is the Bohr magneton, $m_\nu$ is the neutrino mass and $e$ is the electric charge.
Taking into account the current upper limit on the neutrino mass of the order of 1~eV, this value is about 8 orders of magnitude smaller than the Borexino and GEMMA limits.
However,
searching for values of $\mu_{\nu}$ larger than that in Eq.~(\ref{ESMmunu})
is interesting because a positive result would represent a clear signal of physics beyond the minimally extended SM (see Ref.~\cite{Giunti:2014ixa}). 

The existence of a neutrino magnetic moment could have a significant effect on the \cenas\ cross-section,
that acquires an additional term
\begin{eqnarray} \nonumber
    \left.
    \frac{d\sigma^{\mathrm{CE}\nu\mathrm{AS}}}{dT_{R}}
    \right|_{\mu_\nu \neq 0}
    &\simeq&
    \frac{d\sigma^{\mathrm{CE}\nu\mathrm{AS}}}{dT_{R}} 
    +
    \frac{\pi \alpha^2 Z^2}{m_e^2}\left(\frac{\mu_{\nu}}{\mu_{B}}\right)^2 \\
    &\cdot&\left(\frac{1}{T_R}-\frac{1}{E_{\nu}}\right)
    (1-F_{\mathrm{e}}(T_R))^2\,,
    \label{magneticmomentcontr}
\end{eqnarray}
where $\alpha$ is the fine structure constant and $m_e$ is the electron mass. The atomic effect is included in the term $ (1-F_{\mathrm{e}}(T_R))^2$. In fact, for high energy neutrinos, the electron form factor goes to zero, obtaining the neutrino magnetic moment contribution for CE$\nu$NS process. On the contrary, for low energies, where $F_{\mathrm{e}}(T_R)\rightarrow 1$, neutrinos see the target as a whole neutral object, thus no electromagnetic properties can affect the process. By adding this contribution in Eq.~(\ref{eq:diff_rate}), integrating for a time period of 5 years, and considering a magnetic moment of $10^{-12}\,\mu_B$, one obtains the expected differential number of \cenas\ events as a function of the atomic recoil energy shown by the blue dotted-dashed line in Fig.~\ref{fig:diff_rate}. Comparing it to the red dashed curve obtained considering a null magnetic moment, it is clear that there is a large difference between the two cases and this difference could be exploited to put stringent limits on the neutrino magnetic moment. 

To estimate the sensitivity, we consider the least-squares function
\begin{equation}
\label{eq:chi2munu}
    \chi^2(\mu_{\nu})=\left(\frac{\mathcal{N}^{\mathrm{CE}\nu\mathrm{AS}}_{\mathrm{SM}}-\mathcal{N}^{\mathrm{CE}\nu\mathrm{AS}}(\mu_{\nu})}{\sqrt{\mathcal{N}^{\mathrm{CE}\nu\mathrm{AS}}_{\mathrm{SM}}}}\right)^2,
\end{equation}
where $\mathcal{N}^{\mathrm{CE}\nu\mathrm{AS}}_{\mathrm{SM}}$ represents the number of \cenas\ events that one would observe if the magnetic moment is zero, while $\mathcal{N}^{\mathrm{CE}\nu\mathrm{AS}}(\mu_{\nu})$ is the number of events for a given value of the magnetic moment. 
\begin{figure}[t!]
\centering
\includegraphics[scale=0.30]{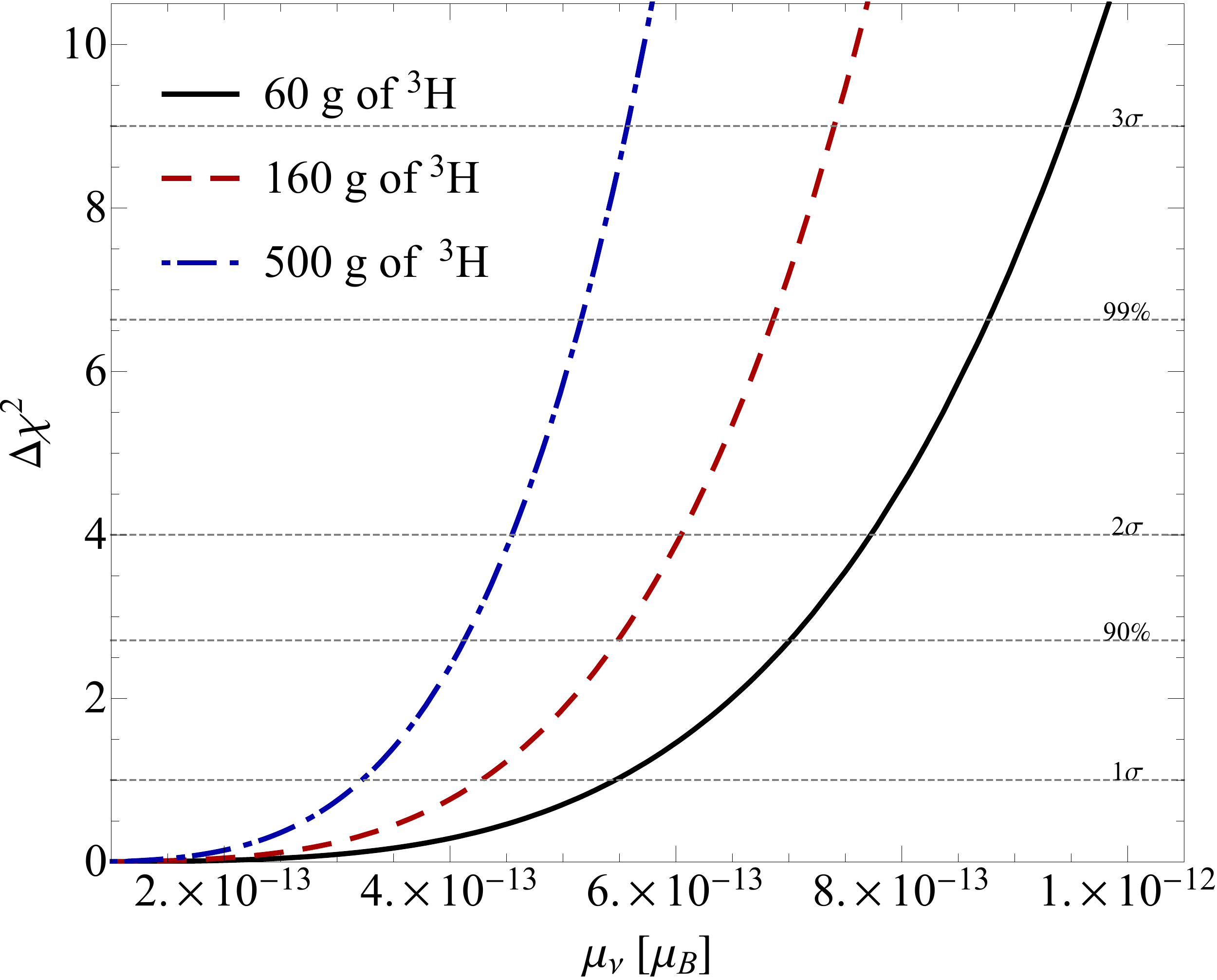}
\caption{With  the  black  solid  curve  it  is  shown  the  \mbox{$\Delta \chi^2 = \chi^2 - \chi^2_{\mathrm{min}}$}, where the $\chi^2$ is defined in Eq.~(\ref{eq:chi2sintheta}), as a function of the magnetic moment of the neutrino $\mu_\nu$, obtained considering a tritium source of 60 g, while with the dashed red line and the dotted-dashed blue line it is shown the $\Delta \chi^2$ obtained considering a tritium source of 160 g and 500 g, respectively. \label{fig:munu}}
\end{figure}
Figure~\ref{fig:munu} shows the $\Delta \chi^2$ profile as a function of the neutrino magnetic moment for the three different experimental scenarios described above: the black solid, red dashed and blue dotted-dashed lines refer to 60 g, 160 g, and 500 g of tritium, respectively. The respective limits that are achievable in these three different scenarios at 90\% C.L. are  $7.0\times10^{-13}\,\mu_B$, $5.5\times10^{-13}\,\mu_B$, and $4.1\times10^{-13}\,\mu_B$, which are almost two orders of magnitude smaller than the current experimental limits.

\section{CONCLUSIONS}

In this paper we proposed an experimental setup to observe coherent elastic neutrino-atom scattering and we evaluate the physics potentialities of this apparatus. The observation of a coherent scattering of the whole atom requires the detection of very low atomic recoil energies, of the order of 10 meV. This is achievable thanks to the combination of different critical ingredients. First, the exploitation of the $\beta^-$ decay of tritium that is characterized by a small $Q$-value,
ensuring a sufficient flux of low-energy antineutrinos. Second, the usage of a target detector with a new technology based on the evaporation of helium atoms coupled with field ionization detector arrays that allows to be sensitive to very small energy deposits. The usage of liquid helium as a target has the triple advantage of being stable, having a small binding energy to the surface, expected to be below 1~meV, and having a small atomic radius $R_{\mathrm{atom}}$, which allows to reach the coherence condition for larger values of the momentum transfer. The effect of the interference between the nucleus and electron cloud scattering cross section is to produce a sharp dip in the recoil spectrum at atomic recoil energies around $T_R\simeq9$~meV, which reduces sizeably the expected number of events with respect to that in the case of coherent neutrino-nucleus elastic scattering. We estimated that in order to observe at 3$\sigma$ the existence of \cenas\ processes it is necessary to use a tritium source of about 60~g, a tank filled with about 500~kg of liquid helium (contained in a cylinder of height 160~cm and radius 90~cm) and a data-taking period of about 5 years.

Keeping the same amount of helium and data-taking period we tested the sensitivity of this experimental setup to the measurement of the Weinberg angle and the observation of a possible neutrino magnetic moment for three different scenarios corresponding to different amounts of tritium in the source: 60~g, 160~g, and 500~g. In the latter scenario, the SM value of the Weinberg angle can be measured with a statistical uncertainty of $\sin^2{\vartheta_W^{\mathrm{SM}}}^{+0.015}_{-0.016}$. Even if this measurement will be less precise than other determinations of the Weinberg angle, it will represent the lowest-energy measurement of $\sin^2{\vartheta_W}$, at a momentum transfer about 2 orders of magnitude smaller than the current determination using caesium atomic parity violation~\cite{Wood:1997zq,Dzuba:2012kx}.
Moreover, the measurement proposed in this paper is not affected by
the uncertainties on the nuclear neutron form factor that 
contribute to atomic parity violation measurements of
$\sin^2{\vartheta_W}$~\cite{Cadeddu:2018izq}.
Therefore,
the observation of \cenas\ would represent an independent test of the
running of $\sin^2{\vartheta_W}$ predicted by the SM
and could be sensitive to low-energy deviations due to physics beyond the SM,
as the contribution of low-mass dark $Z$ bosons.

Finally, we studied the sensitivity of the proposed apparatus to a possible electron neutrino magnetic moment and we found that using 60~g of tritium it is possible to set a lower limit of about $7\times10^{-13}\,\mu_B$ at 90\% C.L., that is more than one order of magnitude smaller than the current experimental limit, showing the great potentialities of this experimental setup.

\section*{Acknowledgment}
M.C., F.D. and E.P. would like to thank Michele Saba and Marco Razeti for the useful suggestions and stimulating discussions about the experimental setup of the detector.

\nocite{*}
\bibliographystyle{apsrev4-1}
\bibliography{bib}

\begin{thebibliography}{57}%
\makeatletter
\providecommand \@ifxundefined [1]{%
 \@ifx{#1\undefined}
}%
\providecommand \@ifnum [1]{%
 \ifnum #1\expandafter \@firstoftwo
 \else \expandafter \@secondoftwo
 \fi
}%
\providecommand \@ifx [1]{%
 \ifx #1\expandafter \@firstoftwo
 \else \expandafter \@secondoftwo
 \fi
}%
\providecommand \natexlab [1]{#1}%
\providecommand \enquote  [1]{``#1''}%
\providecommand \bibnamefont  [1]{#1}%
\providecommand \bibfnamefont [1]{#1}%
\providecommand \citenamefont [1]{#1}%
\providecommand \href@noop [0]{\@secondoftwo}%
\providecommand \href [0]{\begingroup \@sanitize@url \@href}%
\providecommand \@href[1]{\@@startlink{#1}\@@href}%
\providecommand \@@href[1]{\endgroup#1\@@endlink}%
\providecommand \@sanitize@url [0]{\catcode `\\12\catcode `\$12\catcode
  `\&12\catcode `\#12\catcode `\^12\catcode `\_12\catcode `\%12\relax}%
\providecommand \@@startlink[1]{}%
\providecommand \@@endlink[0]{}%
\providecommand \url  [0]{\begingroup\@sanitize@url \@url }%
\providecommand \@url [1]{\endgroup\@href {#1}{\urlprefix }}%
\providecommand \urlprefix  [0]{URL }%
\providecommand \Eprint [0]{\href }%
\providecommand \doibase [0]{http://dx.doi.org/}%
\providecommand \selectlanguage [0]{\@gobble}%
\providecommand \bibinfo  [0]{\@secondoftwo}%
\providecommand \bibfield  [0]{\@secondoftwo}%
\providecommand \translation [1]{[#1]}%
\providecommand \BibitemOpen [0]{}%
\providecommand \bibitemStop [0]{}%
\providecommand \bibitemNoStop [0]{.\EOS\space}%
\providecommand \EOS [0]{\spacefactor3000\relax}%
\providecommand \BibitemShut  [1]{\csname bibitem#1\endcsname}%
\let\auto@bib@innerbib\@empty
\bibitem [{\citenamefont {Akimov}\ \emph {et~al.}(2017)\citenamefont {Akimov}
  \emph {et~al.}}]{Akimov:2017ade}%
  \BibitemOpen
  \bibfield  {author} {\bibinfo {author} {\bibfnamefont {D.}~\bibnamefont
  {Akimov}} \emph {et~al.} (\bibinfo {collaboration} {COHERENT}),\ }\href
  {\doibase 10.1126/science.aao0990} {\bibfield  {journal} {\bibinfo  {journal}
  {Science}\ }\textbf {\bibinfo {volume} {357}},\ \bibinfo {pages} {1123}
  (\bibinfo {year} {2017})},\ \Eprint {http://arxiv.org/abs/1708.01294}
  {arXiv:1708.01294 [nucl-ex]} \BibitemShut {NoStop}%
\bibitem [{\citenamefont {Akimov}\ \emph {et~al.}(2018)\citenamefont {Akimov}
  \emph {et~al.}}]{Akimov:2018vzs}%
  \BibitemOpen
  \bibfield  {author} {\bibinfo {author} {\bibfnamefont {D.}~\bibnamefont
  {Akimov}} \emph {et~al.} (\bibinfo {collaboration} {COHERENT}),\ }\href
  {\doibase 10.5281/zenodo.1228631} {\  (\bibinfo {year} {2018}),\
  10.5281/zenodo.1228631},\ \Eprint {http://arxiv.org/abs/1804.09459}
  {arXiv:1804.09459 [nucl-ex]} \BibitemShut {NoStop}%
\bibitem [{\citenamefont {Freedman}(1974)}]{PhysRevD.9.1389}%
  \BibitemOpen
  \bibfield  {author} {\bibinfo {author} {\bibfnamefont {D.~Z.}\ \bibnamefont
  {Freedman}},\ }\href {\doibase 10.1103/PhysRevD.9.1389} {\bibfield  {journal}
  {\bibinfo  {journal} {Phys. Rev. D}\ }\textbf {\bibinfo {volume} {9}},\
  \bibinfo {pages} {1389} (\bibinfo {year} {1974})}\BibitemShut {NoStop}%
\bibitem [{\citenamefont {Freedman}\ \emph {et~al.}(1977)\citenamefont
  {Freedman}, \citenamefont {Schramm},\ and\ \citenamefont
  {Tubbs}}]{Freedman:1977xn}%
  \BibitemOpen
  \bibfield  {author} {\bibinfo {author} {\bibfnamefont {D.~Z.}\ \bibnamefont
  {Freedman}}, \bibinfo {author} {\bibfnamefont {D.~N.}\ \bibnamefont
  {Schramm}}, \ and\ \bibinfo {author} {\bibfnamefont {D.~L.}\ \bibnamefont
  {Tubbs}},\ }\href {\doibase 10.1146/annurev.ns.27.120177.001123} {\bibfield
  {journal} {\bibinfo  {journal} {Ann. Rev. Nucl. Part. Sci.}\ }\textbf
  {\bibinfo {volume} {27}},\ \bibinfo {pages} {167} (\bibinfo {year}
  {1977})}\BibitemShut {NoStop}%
\bibitem [{\citenamefont {Drukier}\ and\ \citenamefont
  {Stodolsky}(1984)}]{Drukier:1983gj}%
  \BibitemOpen
  \bibfield  {author} {\bibinfo {author} {\bibfnamefont {A.}~\bibnamefont
  {Drukier}}\ and\ \bibinfo {author} {\bibfnamefont {L.}~\bibnamefont
  {Stodolsky}},\ }\href {\doibase 10.1103/PhysRevD.30.2295} {\bibfield
  {journal} {\bibinfo  {journal} {Phys. Rev.}\ }\textbf {\bibinfo {volume}
  {D30}},\ \bibinfo {pages} {2295} (\bibinfo {year} {1984})},\ \bibinfo {note}
  {[,395(1984)]}\BibitemShut {NoStop}%
\bibitem [{\citenamefont {Cadeddu}\ \emph
  {et~al.}(2018{\natexlab{a}})\citenamefont {Cadeddu}, \citenamefont {Giunti},
  \citenamefont {Li},\ and\ \citenamefont {Zhang}}]{Cadeddu:2017etk}%
  \BibitemOpen
  \bibfield  {author} {\bibinfo {author} {\bibfnamefont {M.}~\bibnamefont
  {Cadeddu}}, \bibinfo {author} {\bibfnamefont {C.}~\bibnamefont {Giunti}},
  \bibinfo {author} {\bibfnamefont {Y.~F.}\ \bibnamefont {Li}}, \ and\ \bibinfo
  {author} {\bibfnamefont {Y.~Y.}\ \bibnamefont {Zhang}},\ }\href {\doibase
  10.1103/PhysRevLett.120.072501} {\bibfield  {journal} {\bibinfo  {journal}
  {Phys. Rev. Lett.}\ }\textbf {\bibinfo {volume} {120}},\ \bibinfo {pages}
  {072501} (\bibinfo {year} {2018}{\natexlab{a}})},\ \Eprint
  {http://arxiv.org/abs/1710.02730} {arXiv:1710.02730 [hep-ph]} \BibitemShut
  {NoStop}%
\bibitem [{\citenamefont {Papoulias}\ \emph {et~al.}(2019)\citenamefont
  {Papoulias}, \citenamefont {Kosmas}, \citenamefont {Sahu}, \citenamefont
  {Kota},\ and\ \citenamefont {Hota}}]{Papoulias:2019lfi}%
  \BibitemOpen
  \bibfield  {author} {\bibinfo {author} {\bibfnamefont {D.~K.}\ \bibnamefont
  {Papoulias}}, \bibinfo {author} {\bibfnamefont {T.~S.}\ \bibnamefont
  {Kosmas}}, \bibinfo {author} {\bibfnamefont {R.}~\bibnamefont {Sahu}},
  \bibinfo {author} {\bibfnamefont {V.~K.~B.}\ \bibnamefont {Kota}}, \ and\
  \bibinfo {author} {\bibfnamefont {M.}~\bibnamefont {Hota}},\ }\href@noop {}
  {\  (\bibinfo {year} {2019})},\ \Eprint {http://arxiv.org/abs/1903.03722}
  {arXiv:1903.03722 [hep-ph]} \BibitemShut {NoStop}%
\bibitem [{\citenamefont {Coloma}\ \emph {et~al.}(2017)\citenamefont {Coloma},
  \citenamefont {Gonzalez-Garcia}, \citenamefont {Maltoni},\ and\ \citenamefont
  {Schwetz}}]{Coloma:2017ncl}%
  \BibitemOpen
  \bibfield  {author} {\bibinfo {author} {\bibfnamefont {P.}~\bibnamefont
  {Coloma}}, \bibinfo {author} {\bibfnamefont {M.~C.}\ \bibnamefont
  {Gonzalez-Garcia}}, \bibinfo {author} {\bibfnamefont {M.}~\bibnamefont
  {Maltoni}}, \ and\ \bibinfo {author} {\bibfnamefont {T.}~\bibnamefont
  {Schwetz}},\ }\href {\doibase 10.1103/PhysRevD.96.115007} {\bibfield
  {journal} {\bibinfo  {journal} {Phys. Rev.}\ }\textbf {\bibinfo {volume}
  {D96}},\ \bibinfo {pages} {115007} (\bibinfo {year} {2017})},\ \Eprint
  {http://arxiv.org/abs/1708.02899} {arXiv:1708.02899 [hep-ph]} \BibitemShut
  {NoStop}%
\bibitem [{\citenamefont {Miranda}\ \emph
  {et~al.}(2019{\natexlab{a}})\citenamefont {Miranda}, \citenamefont
  {Papoulias}, \citenamefont {Tórtola},\ and\ \citenamefont
  {Valle}}]{Miranda:2019wdy}%
  \BibitemOpen
  \bibfield  {author} {\bibinfo {author} {\bibfnamefont {O.~G.}\ \bibnamefont
  {Miranda}}, \bibinfo {author} {\bibfnamefont {D.~K.}\ \bibnamefont
  {Papoulias}}, \bibinfo {author} {\bibfnamefont {M.}~\bibnamefont {Tórtola}},
  \ and\ \bibinfo {author} {\bibfnamefont {J.~W.~F.}\ \bibnamefont {Valle}},\
  }\href {\doibase 10.1007/JHEP07(2019)103} {\bibfield  {journal} {\bibinfo
  {journal} {JHEP}\ }\textbf {\bibinfo {volume} {07}},\ \bibinfo {pages} {103}
  (\bibinfo {year} {2019}{\natexlab{a}})},\ \Eprint
  {http://arxiv.org/abs/1905.03750} {arXiv:1905.03750 [hep-ph]} \BibitemShut
  {NoStop}%
\bibitem [{\citenamefont {Cadeddu}\ \emph
  {et~al.}(2018{\natexlab{b}})\citenamefont {Cadeddu}, \citenamefont {Giunti},
  \citenamefont {Kouzakov}, \citenamefont {Li}, \citenamefont {Studenikin},\
  and\ \citenamefont {Zhang}}]{Cadeddu:2018dux}%
  \BibitemOpen
  \bibfield  {author} {\bibinfo {author} {\bibfnamefont {M.}~\bibnamefont
  {Cadeddu}}, \bibinfo {author} {\bibfnamefont {C.}~\bibnamefont {Giunti}},
  \bibinfo {author} {\bibfnamefont {K.~A.}\ \bibnamefont {Kouzakov}}, \bibinfo
  {author} {\bibfnamefont {Y.~F.}\ \bibnamefont {Li}}, \bibinfo {author}
  {\bibfnamefont {A.~I.}\ \bibnamefont {Studenikin}}, \ and\ \bibinfo {author}
  {\bibfnamefont {Y.~Y.}\ \bibnamefont {Zhang}},\ }\href {\doibase
  10.1103/PhysRevD.98.113010} {\bibfield  {journal} {\bibinfo  {journal} {Phys.
  Rev.}\ }\textbf {\bibinfo {volume} {D98}},\ \bibinfo {pages} {113010}
  (\bibinfo {year} {2018}{\natexlab{b}})},\ \Eprint
  {http://arxiv.org/abs/1810.05606} {arXiv:1810.05606 [hep-ph]} \BibitemShut
  {NoStop}%
\bibitem [{\citenamefont {Dutta}\ \emph {et~al.}(2019)\citenamefont {Dutta},
  \citenamefont {Liao}, \citenamefont {Sinha},\ and\ \citenamefont
  {Strigari}}]{Dutta:2019eml}%
  \BibitemOpen
  \bibfield  {author} {\bibinfo {author} {\bibfnamefont {B.}~\bibnamefont
  {Dutta}}, \bibinfo {author} {\bibfnamefont {S.}~\bibnamefont {Liao}},
  \bibinfo {author} {\bibfnamefont {S.}~\bibnamefont {Sinha}}, \ and\ \bibinfo
  {author} {\bibfnamefont {L.~E.}\ \bibnamefont {Strigari}},\ }\href {\doibase
  10.1103/PhysRevLett.123.061801} {\bibfield  {journal} {\bibinfo  {journal}
  {Phys. Rev. Lett.}\ }\textbf {\bibinfo {volume} {123}},\ \bibinfo {pages}
  {061801} (\bibinfo {year} {2019})},\ \Eprint
  {http://arxiv.org/abs/1903.10666} {arXiv:1903.10666 [hep-ph]} \BibitemShut
  {NoStop}%
\bibitem [{\citenamefont {Miranda}\ \emph
  {et~al.}(2019{\natexlab{b}})\citenamefont {Miranda}, \citenamefont
  {Sanchez~Garcia},\ and\ \citenamefont {Sanders}}]{Miranda:2019skf}%
  \BibitemOpen
  \bibfield  {author} {\bibinfo {author} {\bibfnamefont {O.~G.}\ \bibnamefont
  {Miranda}}, \bibinfo {author} {\bibfnamefont {G.}~\bibnamefont
  {Sanchez~Garcia}}, \ and\ \bibinfo {author} {\bibfnamefont {O.}~\bibnamefont
  {Sanders}},\ }\href {\doibase 10.1155/2019/3902819} {\bibfield  {journal}
  {\bibinfo  {journal} {Adv. High Energy Phys.}\ }\textbf {\bibinfo {volume}
  {2019}},\ \bibinfo {pages} {3902819} (\bibinfo {year}
  {2019}{\natexlab{b}})},\ \Eprint {http://arxiv.org/abs/1902.09036}
  {arXiv:1902.09036 [hep-ph]} \BibitemShut {NoStop}%
\bibitem [{\citenamefont {Heeck}\ \emph {et~al.}(2019)\citenamefont {Heeck},
  \citenamefont {Lindner}, \citenamefont {Rodejohann},\ and\ \citenamefont
  {Vogl}}]{Heeck:2018nzc}%
  \BibitemOpen
  \bibfield  {author} {\bibinfo {author} {\bibfnamefont {J.}~\bibnamefont
  {Heeck}}, \bibinfo {author} {\bibfnamefont {M.}~\bibnamefont {Lindner}},
  \bibinfo {author} {\bibfnamefont {W.}~\bibnamefont {Rodejohann}}, \ and\
  \bibinfo {author} {\bibfnamefont {S.}~\bibnamefont {Vogl}},\ }\href {\doibase
  10.21468/SciPostPhys.6.3.038} {\bibfield  {journal} {\bibinfo  {journal}
  {SciPost Phys.}\ }\textbf {\bibinfo {volume} {6}},\ \bibinfo {pages} {038}
  (\bibinfo {year} {2019})},\ \Eprint {http://arxiv.org/abs/1812.04067}
  {arXiv:1812.04067 [hep-ph]} \BibitemShut {NoStop}%
\bibitem [{\citenamefont {Abdullah}\ \emph {et~al.}(2018)\citenamefont
  {Abdullah}, \citenamefont {Dent}, \citenamefont {Dutta}, \citenamefont
  {Kane}, \citenamefont {Liao},\ and\ \citenamefont
  {Strigari}}]{Abdullah:2018ykz}%
  \BibitemOpen
  \bibfield  {author} {\bibinfo {author} {\bibfnamefont {M.}~\bibnamefont
  {Abdullah}}, \bibinfo {author} {\bibfnamefont {J.~B.}\ \bibnamefont {Dent}},
  \bibinfo {author} {\bibfnamefont {B.}~\bibnamefont {Dutta}}, \bibinfo
  {author} {\bibfnamefont {G.~L.}\ \bibnamefont {Kane}}, \bibinfo {author}
  {\bibfnamefont {S.}~\bibnamefont {Liao}}, \ and\ \bibinfo {author}
  {\bibfnamefont {L.~E.}\ \bibnamefont {Strigari}},\ }\href {\doibase
  10.1103/PhysRevD.98.015005} {\bibfield  {journal} {\bibinfo  {journal} {Phys.
  Rev.}\ }\textbf {\bibinfo {volume} {D98}},\ \bibinfo {pages} {015005}
  (\bibinfo {year} {2018})},\ \Eprint {http://arxiv.org/abs/1803.01224}
  {arXiv:1803.01224 [hep-ph]} \BibitemShut {NoStop}%
\bibitem [{\citenamefont {Farzan}\ \emph {et~al.}(2018)\citenamefont {Farzan},
  \citenamefont {Lindner}, \citenamefont {Rodejohann},\ and\ \citenamefont
  {Xu}}]{Farzan:2018gtr}%
  \BibitemOpen
  \bibfield  {author} {\bibinfo {author} {\bibfnamefont {Y.}~\bibnamefont
  {Farzan}}, \bibinfo {author} {\bibfnamefont {M.}~\bibnamefont {Lindner}},
  \bibinfo {author} {\bibfnamefont {W.}~\bibnamefont {Rodejohann}}, \ and\
  \bibinfo {author} {\bibfnamefont {X.-J.}\ \bibnamefont {Xu}},\ }\href
  {\doibase 10.1007/JHEP05(2018)066} {\bibfield  {journal} {\bibinfo  {journal}
  {JHEP}\ }\textbf {\bibinfo {volume} {05}},\ \bibinfo {pages} {066} (\bibinfo
  {year} {2018})},\ \Eprint {http://arxiv.org/abs/1802.05171} {arXiv:1802.05171
  [hep-ph]} \BibitemShut {NoStop}%
\bibitem [{\citenamefont {Liao}\ and\ \citenamefont
  {Marfatia}(2017)}]{Liao:2017uzy}%
  \BibitemOpen
  \bibfield  {author} {\bibinfo {author} {\bibfnamefont {J.}~\bibnamefont
  {Liao}}\ and\ \bibinfo {author} {\bibfnamefont {D.}~\bibnamefont
  {Marfatia}},\ }\href {\doibase 10.1016/j.physletb.2017.10.046} {\bibfield
  {journal} {\bibinfo  {journal} {Phys. Lett.}\ }\textbf {\bibinfo {volume}
  {B775}},\ \bibinfo {pages} {54} (\bibinfo {year} {2017})},\ \Eprint
  {http://arxiv.org/abs/1708.04255} {arXiv:1708.04255 [hep-ph]} \BibitemShut
  {NoStop}%
\bibitem [{\citenamefont {Papoulias}\ and\ \citenamefont
  {Kosmas}(2018)}]{Kosmas:2017tsq}%
  \BibitemOpen
  \bibfield  {author} {\bibinfo {author} {\bibfnamefont {D.~K.}\ \bibnamefont
  {Papoulias}}\ and\ \bibinfo {author} {\bibfnamefont {T.~S.}\ \bibnamefont
  {Kosmas}},\ }\href {\doibase 10.1103/PhysRevD.97.033003} {\bibfield
  {journal} {\bibinfo  {journal} {Phys. Rev.}\ }\textbf {\bibinfo {volume}
  {D97}},\ \bibinfo {pages} {033003} (\bibinfo {year} {2018})},\ \Eprint
  {http://arxiv.org/abs/1711.09773} {arXiv:1711.09773 [hep-ph]} \BibitemShut
  {NoStop}%
\bibitem [{\citenamefont {Cadeddu}\ and\ \citenamefont
  {Dordei}(2019)}]{Cadeddu:2018izq}%
  \BibitemOpen
  \bibfield  {author} {\bibinfo {author} {\bibfnamefont {M.}~\bibnamefont
  {Cadeddu}}\ and\ \bibinfo {author} {\bibfnamefont {F.}~\bibnamefont
  {Dordei}},\ }\href {\doibase 10.1103/PhysRevD.99.033010} {\bibfield
  {journal} {\bibinfo  {journal} {Phys. Rev.}\ }\textbf {\bibinfo {volume}
  {D99}},\ \bibinfo {pages} {033010} (\bibinfo {year} {2019})},\ \Eprint
  {http://arxiv.org/abs/1808.10202} {arXiv:1808.10202 [hep-ph]} \BibitemShut
  {NoStop}%
\bibitem [{\citenamefont {Cañas}\ \emph {et~al.}(2018)\citenamefont {Cañas},
  \citenamefont {Garcés}, \citenamefont {Miranda},\ and\ \citenamefont
  {Parada}}]{Canas:2018rng}%
  \BibitemOpen
  \bibfield  {author} {\bibinfo {author} {\bibfnamefont {B.~C.}\ \bibnamefont
  {Cañas}}, \bibinfo {author} {\bibfnamefont {E.~A.}\ \bibnamefont {Garcés}},
  \bibinfo {author} {\bibfnamefont {O.~G.}\ \bibnamefont {Miranda}}, \ and\
  \bibinfo {author} {\bibfnamefont {A.}~\bibnamefont {Parada}},\ }\href
  {\doibase 10.1016/j.physletb.2018.07.049} {\bibfield  {journal} {\bibinfo
  {journal} {Phys. Lett.}\ }\textbf {\bibinfo {volume} {B784}},\ \bibinfo
  {pages} {159} (\bibinfo {year} {2018})},\ \Eprint
  {http://arxiv.org/abs/1806.01310} {arXiv:1806.01310 [hep-ph]} \BibitemShut
  {NoStop}%
\bibitem [{\citenamefont {Bednyakov}\ and\ \citenamefont
  {Naumov}(2018)}]{Bednyakov:2018mjd}%
  \BibitemOpen
  \bibfield  {author} {\bibinfo {author} {\bibfnamefont {V.~A.}\ \bibnamefont
  {Bednyakov}}\ and\ \bibinfo {author} {\bibfnamefont {D.~V.}\ \bibnamefont
  {Naumov}},\ }\href {\doibase 10.1103/PhysRevD.98.053004} {\bibfield
  {journal} {\bibinfo  {journal} {Phys. Rev.}\ }\textbf {\bibinfo {volume}
  {D98}},\ \bibinfo {pages} {053004} (\bibinfo {year} {2018})},\ \Eprint
  {http://arxiv.org/abs/1806.08768} {arXiv:1806.08768 [hep-ph]} \BibitemShut
  {NoStop}%
\bibitem [{\citenamefont {Sehgal}\ and\ \citenamefont
  {Wanninger}(1986)}]{Sehgal1986}%
  \BibitemOpen
  \bibfield  {author} {\bibinfo {author} {\bibfnamefont {L.~M.}\ \bibnamefont
  {Sehgal}}\ and\ \bibinfo {author} {\bibfnamefont {M.}~\bibnamefont
  {Wanninger}},\ }\href {\doibase 10.1016/0370-2693(86)91008-7} {\bibfield
  {journal} {\bibinfo  {journal} {Physics Letters B}\ }\textbf {\bibinfo
  {volume} {171}},\ \bibinfo {pages} {107} (\bibinfo {year}
  {1986})}\BibitemShut {NoStop}%
\bibitem [{\citenamefont {Maris}\ \emph {et~al.}(2017)\citenamefont {Maris},
  \citenamefont {Seidel},\ and\ \citenamefont {Stein}}]{Maris:2017xvi}%
  \BibitemOpen
  \bibfield  {author} {\bibinfo {author} {\bibfnamefont {H.~J.}\ \bibnamefont
  {Maris}}, \bibinfo {author} {\bibfnamefont {G.~M.}\ \bibnamefont {Seidel}}, \
  and\ \bibinfo {author} {\bibfnamefont {D.}~\bibnamefont {Stein}},\ }\href
  {\doibase 10.1103/PhysRevLett.119.181303} {\bibfield  {journal} {\bibinfo
  {journal} {Phys. Rev. Lett.}\ }\textbf {\bibinfo {volume} {119}},\ \bibinfo
  {pages} {181303} (\bibinfo {year} {2017})},\ \Eprint
  {http://arxiv.org/abs/1706.00117} {arXiv:1706.00117 [astro-ph.IM]}
  \BibitemShut {NoStop}%
\bibitem [{\citenamefont {Betts}\ \emph {et~al.}(2013)\citenamefont {Betts}
  \emph {et~al.}}]{Betts:2013uya}%
  \BibitemOpen
  \bibfield  {author} {\bibinfo {author} {\bibfnamefont {S.}~\bibnamefont
  {Betts}} \emph {et~al.},\ }in\ \href
  {http://www.slac.stanford.edu/econf/C1307292/docs/submittedArxivFiles/1307.4738.pdf}
  {\emph {\bibinfo {booktitle} {{Proceedings, 2013 Community Summer Study on
  the Future of U.S. Particle Physics: Snowmass on the Mississippi (CSS2013):
  Minneapolis, MN, USA, July 29-August 6, 2013}}}}\ (\bibinfo {year} {2013})\
  \Eprint {http://arxiv.org/abs/1307.4738} {arXiv:1307.4738 [astro-ph.IM]}
  \BibitemShut {NoStop}%
\bibitem [{\citenamefont {Betti}\ \emph {et~al.}(2019)\citenamefont {Betti}
  \emph {et~al.}}]{Betti:2019ouf}%
  \BibitemOpen
  \bibfield  {author} {\bibinfo {author} {\bibfnamefont {M.~G.}\ \bibnamefont
  {Betti}} \emph {et~al.} (\bibinfo {collaboration} {PTOLEMY}),\ }\href
  {\doibase 10.1088/1475-7516/2019/07/047} {\bibfield  {journal} {\bibinfo
  {journal} {JCAP}\ }\textbf {\bibinfo {volume} {1907}},\ \bibinfo {pages}
  {047} (\bibinfo {year} {2019})},\ \Eprint {http://arxiv.org/abs/1902.05508}
  {arXiv:1902.05508 [astro-ph.CO]} \BibitemShut {NoStop}%
\bibitem [{\citenamefont {Gaponov}\ and\ \citenamefont
  {Tikhonov}(1977)}]{gaponov}%
  \BibitemOpen
  \bibfield  {author} {\bibinfo {author} {\bibfnamefont {Y.~V.}\ \bibnamefont
  {Gaponov}}\ and\ \bibinfo {author} {\bibfnamefont {V.}~\bibnamefont
  {Tikhonov}},\ }\href@noop {} {\bibfield  {journal} {\bibinfo  {journal}
  {Yadernaya Fizika}\ }\textbf {\bibinfo {volume} {26}},\ \bibinfo {pages}
  {594} (\bibinfo {year} {1977})}\BibitemShut {NoStop}%
\bibitem [{\citenamefont {Patton}\ \emph {et~al.}(2012)\citenamefont {Patton},
  \citenamefont {Engel}, \citenamefont {McLaughlin},\ and\ \citenamefont
  {Schunck}}]{PhysRevC.86.024612}%
  \BibitemOpen
  \bibfield  {author} {\bibinfo {author} {\bibfnamefont {K.}~\bibnamefont
  {Patton}}, \bibinfo {author} {\bibfnamefont {J.}~\bibnamefont {Engel}},
  \bibinfo {author} {\bibfnamefont {G.~C.}\ \bibnamefont {McLaughlin}}, \ and\
  \bibinfo {author} {\bibfnamefont {N.}~\bibnamefont {Schunck}},\ }\href
  {\doibase 10.1103/PhysRevC.86.024612} {\bibfield  {journal} {\bibinfo
  {journal} {Phys. Rev. C}\ }\textbf {\bibinfo {volume} {86}},\ \bibinfo
  {pages} {024612} (\bibinfo {year} {2012})}\BibitemShut {NoStop}%
\bibitem [{\citenamefont {Barranco}\ \emph {et~al.}(2005)\citenamefont
  {Barranco}, \citenamefont {Miranda},\ and\ \citenamefont
  {Rashba}}]{Barranco:2005yy}%
  \BibitemOpen
  \bibfield  {author} {\bibinfo {author} {\bibfnamefont {J.}~\bibnamefont
  {Barranco}}, \bibinfo {author} {\bibfnamefont {O.~G.}\ \bibnamefont
  {Miranda}}, \ and\ \bibinfo {author} {\bibfnamefont {T.~I.}\ \bibnamefont
  {Rashba}},\ }\href {\doibase 10.1088/1126-6708/2005/12/021} {\bibfield
  {journal} {\bibinfo  {journal} {JHEP}\ }\textbf {\bibinfo {volume} {12}},\
  \bibinfo {pages} {021} (\bibinfo {year} {2005})},\ \Eprint
  {http://arxiv.org/abs/hep-ph/0508299} {arXiv:hep-ph/0508299 [hep-ph]}
  \BibitemShut {NoStop}%
\bibitem [{\citenamefont {Aristizabal~Sierra}\ \emph
  {et~al.}(2019)\citenamefont {Aristizabal~Sierra}, \citenamefont {Liao},\ and\
  \citenamefont {Marfatia}}]{AristizabalSierra:2019zmy}%
  \BibitemOpen
  \bibfield  {author} {\bibinfo {author} {\bibfnamefont {D.}~\bibnamefont
  {Aristizabal~Sierra}}, \bibinfo {author} {\bibfnamefont {J.}~\bibnamefont
  {Liao}}, \ and\ \bibinfo {author} {\bibfnamefont {D.}~\bibnamefont
  {Marfatia}},\ }\href {\doibase 10.1007/JHEP06(2019)141} {\bibfield  {journal}
  {\bibinfo  {journal} {JHEP}\ }\textbf {\bibinfo {volume} {06}},\ \bibinfo
  {pages} {141} (\bibinfo {year} {2019})},\ \Eprint
  {http://arxiv.org/abs/1902.07398} {arXiv:1902.07398 [hep-ph]} \BibitemShut
  {NoStop}%
\bibitem [{\citenamefont {Doyle}\ and\ \citenamefont
  {Turner}(1968)}]{Doyle1968}%
  \BibitemOpen
  \bibfield  {author} {\bibinfo {author} {\bibfnamefont {P.~A.}\ \bibnamefont
  {Doyle}}\ and\ \bibinfo {author} {\bibfnamefont {P.~S.}\ \bibnamefont
  {Turner}},\ }\href {\doibase 10.1107/S0567739468000756} {\bibfield  {journal}
  {\bibinfo  {journal} {Acta Crystallographica Section A}\ }\textbf {\bibinfo
  {volume} {24}},\ \bibinfo {pages} {390} (\bibinfo {year} {1968})}\BibitemShut
  {NoStop}%
\bibitem [{\citenamefont {Brown}\ \emph {et~al.}(2006)\citenamefont {Brown}
  \emph {et~al.}}]{Brown}%
  \BibitemOpen
  \bibfield  {author} {\bibinfo {author} {\bibfnamefont {P.~J.}\ \bibnamefont
  {Brown}} \emph {et~al.},\ }\href {\doibase 10.1107/97809553602060000600}
  {\bibfield  {journal} {\bibinfo  {journal} {International Tables for
  Crystallography}\ }\textbf {\bibinfo {volume} {C ch. 6.1}},\ \bibinfo {pages}
  {554} (\bibinfo {year} {2006})}\BibitemShut {NoStop}%
\bibitem [{\citenamefont {Thakkar}\ and\ \citenamefont
  {Smith}(1992)}]{Thakkar1992}%
  \BibitemOpen
  \bibfield  {author} {\bibinfo {author} {\bibfnamefont {A.~J.}\ \bibnamefont
  {Thakkar}}\ and\ \bibinfo {author} {\bibfnamefont {V.~H.}\ \bibnamefont
  {Smith}, \bibfnamefont {Jnr}},\ }\href {\doibase 10.1107/S0108767391005883}
  {\bibfield  {journal} {\bibinfo  {journal} {Acta Crystallographica Section
  A}\ }\textbf {\bibinfo {volume} {48}},\ \bibinfo {pages} {70} (\bibinfo
  {year} {1992})}\BibitemShut {NoStop}%
\bibitem [{\citenamefont {Coulthard}(1967)}]{Coulthard1967}%
  \BibitemOpen
  \bibfield  {author} {\bibinfo {author} {\bibfnamefont {M.~A.}\ \bibnamefont
  {Coulthard}},\ }\href {\doibase 10.1088/0370-1328/91/1/309} {\bibfield
  {journal} {\bibinfo  {journal} {Proceedings of the Physical Society}\
  }\textbf {\bibinfo {volume} {91}},\ \bibinfo {pages} {44} (\bibinfo {year}
  {1967})}\BibitemShut {NoStop}%
\bibitem [{\citenamefont {Thakkar}\ and\ \citenamefont
  {Smith}(1977)}]{Thakkar1977}%
  \BibitemOpen
  \bibfield  {author} {\bibinfo {author} {\bibfnamefont {A.~J.}\ \bibnamefont
  {Thakkar}}\ and\ \bibinfo {author} {\bibfnamefont {V.~H.}\ \bibnamefont
  {Smith}},\ }\href {\doibase 10.1103/PhysRevA.15.1} {\bibfield  {journal}
  {\bibinfo  {journal} {Phys. Rev. A}\ }\textbf {\bibinfo {volume} {15}},\
  \bibinfo {pages} {1} (\bibinfo {year} {1977})}\BibitemShut {NoStop}%
\bibitem [{\citenamefont {Giunti}\ and\ \citenamefont
  {Kim}(2007)}]{Giunti:2007ry}%
  \BibitemOpen
  \bibfield  {author} {\bibinfo {author} {\bibfnamefont {C.}~\bibnamefont
  {Giunti}}\ and\ \bibinfo {author} {\bibfnamefont {C.~W.}\ \bibnamefont
  {Kim}},\ }\href@noop {} {\emph {\bibinfo {title} {{Fundamentals of Neutrino
  Physics and Astrophysics}}}}\ (\bibinfo  {publisher} {Oxford University
  Press},\ \bibinfo {address} {Oxford, UK},\ \bibinfo {year} {2007})\ pp.\
  \bibinfo {pages} {1--728}\BibitemShut {NoStop}%
\bibitem [{\citenamefont {Tanabashi}\ \emph {et~al.}(2018)\citenamefont
  {Tanabashi} \emph {et~al.}}]{Tanabashi:2018oca}%
  \BibitemOpen
  \bibfield  {author} {\bibinfo {author} {\bibfnamefont {M.}~\bibnamefont
  {Tanabashi}} \emph {et~al.} (\bibinfo {collaboration} {Particle Data
  Group}),\ }\href {\doibase 10.1103/PhysRevD.98.030001} {\bibfield  {journal}
  {\bibinfo  {journal} {Phys. Rev.}\ }\textbf {\bibinfo {volume} {D98}},\
  \bibinfo {pages} {030001} (\bibinfo {year} {2018})}\BibitemShut {NoStop}%
\bibitem [{\citenamefont {Cowan}\ \emph {et~al.}(2011)\citenamefont {Cowan},
  \citenamefont {Cranmer}, \citenamefont {Gross},\ and\ \citenamefont
  {Vitells}}]{Cowan:2010js}%
  \BibitemOpen
  \bibfield  {author} {\bibinfo {author} {\bibfnamefont {G.}~\bibnamefont
  {Cowan}}, \bibinfo {author} {\bibfnamefont {K.}~\bibnamefont {Cranmer}},
  \bibinfo {author} {\bibfnamefont {E.}~\bibnamefont {Gross}}, \ and\ \bibinfo
  {author} {\bibfnamefont {O.}~\bibnamefont {Vitells}},\ }\href {\doibase
  10.1140/epjc/s10052-011-1554-0, 10.1140/epjc/s10052-013-2501-z} {\bibfield
  {journal} {\bibinfo  {journal} {Eur. Phys. J.}\ }\textbf {\bibinfo {volume}
  {C71}},\ \bibinfo {pages} {1554} (\bibinfo {year} {2011})},\ \bibinfo {note}
  {[Erratum: Eur. Phys. J.C73,2501(2013)]},\ \Eprint
  {http://arxiv.org/abs/1007.1727} {arXiv:1007.1727 [physics.data-an]}
  \BibitemShut {NoStop}%
\bibitem [{\citenamefont {Hertel}\ \emph {et~al.}(2018)\citenamefont {Hertel},
  \citenamefont {Biekert}, \citenamefont {Lin}, \citenamefont {Velan},\ and\
  \citenamefont {McKinsey}}]{Hertel:2018arXiv}%
  \BibitemOpen
  \bibfield  {author} {\bibinfo {author} {\bibfnamefont {S.~A.}\ \bibnamefont
  {Hertel}}, \bibinfo {author} {\bibfnamefont {A.}~\bibnamefont {Biekert}},
  \bibinfo {author} {\bibfnamefont {J.}~\bibnamefont {Lin}}, \bibinfo {author}
  {\bibfnamefont {V.}~\bibnamefont {Velan}}, \ and\ \bibinfo {author}
  {\bibfnamefont {D.~N.}\ \bibnamefont {McKinsey}},\ }\href@noop {} {\
  (\bibinfo {year} {2018})},\ \Eprint {http://arxiv.org/abs/1810.06283}
  {arXiv:1810.06283 [ins-det]} \BibitemShut {NoStop}%
\bibitem [{\citenamefont {Wood}\ \emph {et~al.}(1997)\citenamefont {Wood},
  \citenamefont {Bennett}, \citenamefont {Cho}, \citenamefont {Masterson},
  \citenamefont {Roberts}, \citenamefont {Tanner},\ and\ \citenamefont
  {Wieman}}]{Wood:1997zq}%
  \BibitemOpen
  \bibfield  {author} {\bibinfo {author} {\bibfnamefont {C.~S.}\ \bibnamefont
  {Wood}}, \bibinfo {author} {\bibfnamefont {S.~C.}\ \bibnamefont {Bennett}},
  \bibinfo {author} {\bibfnamefont {D.}~\bibnamefont {Cho}}, \bibinfo {author}
  {\bibfnamefont {B.~P.}\ \bibnamefont {Masterson}}, \bibinfo {author}
  {\bibfnamefont {J.~L.}\ \bibnamefont {Roberts}}, \bibinfo {author}
  {\bibfnamefont {C.~E.}\ \bibnamefont {Tanner}}, \ and\ \bibinfo {author}
  {\bibfnamefont {C.~E.}\ \bibnamefont {Wieman}},\ }\href {\doibase
  10.1126/science.275.5307.1759} {\bibfield  {journal} {\bibinfo  {journal}
  {Science}\ }\textbf {\bibinfo {volume} {275}},\ \bibinfo {pages} {1759}
  (\bibinfo {year} {1997})}\BibitemShut {NoStop}%
\bibitem [{\citenamefont {Dzuba}\ \emph {et~al.}(2012)\citenamefont {Dzuba},
  \citenamefont {Berengut}, \citenamefont {Flambaum},\ and\ \citenamefont
  {Roberts}}]{Dzuba:2012kx}%
  \BibitemOpen
  \bibfield  {author} {\bibinfo {author} {\bibfnamefont {V.~A.}\ \bibnamefont
  {Dzuba}}, \bibinfo {author} {\bibfnamefont {J.~C.}\ \bibnamefont {Berengut}},
  \bibinfo {author} {\bibfnamefont {V.~V.}\ \bibnamefont {Flambaum}}, \ and\
  \bibinfo {author} {\bibfnamefont {B.}~\bibnamefont {Roberts}},\ }\href
  {\doibase 10.1103/PhysRevLett.109.203003} {\bibfield  {journal} {\bibinfo
  {journal} {Phys. Rev. Lett.}\ }\textbf {\bibinfo {volume} {109}},\ \bibinfo
  {pages} {203003} (\bibinfo {year} {2012})},\ \Eprint
  {http://arxiv.org/abs/1207.5864} {arXiv:1207.5864 [hep-ph]} \BibitemShut
  {NoStop}%
\bibitem [{\citenamefont {Anthony}\ \emph {et~al.}(2005)\citenamefont {Anthony}
  \emph {et~al.}}]{Anthony:2005pm}%
  \BibitemOpen
  \bibfield  {author} {\bibinfo {author} {\bibfnamefont {P.~L.}\ \bibnamefont
  {Anthony}} \emph {et~al.} (\bibinfo {collaboration} {SLAC E158}),\ }\href
  {\doibase 10.1103/PhysRevLett.95.081601} {\bibfield  {journal} {\bibinfo
  {journal} {Phys. Rev. Lett.}\ }\textbf {\bibinfo {volume} {95}},\ \bibinfo
  {pages} {081601} (\bibinfo {year} {2005})},\ \Eprint
  {http://arxiv.org/abs/hep-ex/0504049} {arXiv:hep-ex/0504049 [hep-ex]}
  \BibitemShut {NoStop}%
\bibitem [{\citenamefont {Wang}\ \emph {et~al.}(2014)\citenamefont {Wang} \emph
  {et~al.}}]{Wang:2014bbc}%
  \BibitemOpen
  \bibfield  {author} {\bibinfo {author} {\bibfnamefont {D.}~\bibnamefont
  {Wang}} \emph {et~al.} (\bibinfo {collaboration} {PVDIS}),\ }\href {\doibase
  10.1038/nature12964} {\bibfield  {journal} {\bibinfo  {journal} {Nature}\
  }\textbf {\bibinfo {volume} {506}},\ \bibinfo {pages} {67} (\bibinfo {year}
  {2014})}\BibitemShut {NoStop}%
\bibitem [{\citenamefont {Zeller}\ \emph {et~al.}(2002)\citenamefont {Zeller}
  \emph {et~al.}}]{Zeller:2001hh}%
  \BibitemOpen
  \bibfield  {author} {\bibinfo {author} {\bibfnamefont {G.~P.}\ \bibnamefont
  {Zeller}} \emph {et~al.} (\bibinfo {collaboration} {NuTeV}),\ }\href
  {\doibase 10.1103/PhysRevLett.88.091802, 10.1103/PhysRevLett.90.239902}
  {\bibfield  {journal} {\bibinfo  {journal} {Phys. Rev. Lett.}\ }\textbf
  {\bibinfo {volume} {88}},\ \bibinfo {pages} {091802} (\bibinfo {year}
  {2002})},\ \bibinfo {note} {[Erratum: Phys. Rev. Lett.90,239902(2003)]},\
  \Eprint {http://arxiv.org/abs/hep-ex/0110059} {arXiv:hep-ex/0110059 [hep-ex]}
  \BibitemShut {NoStop}%
\bibitem [{\citenamefont {Androić}\ \emph {et~al.}(2018)\citenamefont
  {Androić} \emph {et~al.}}]{Androic:2018kni}%
  \BibitemOpen
  \bibfield  {author} {\bibinfo {author} {\bibfnamefont {D.}~\bibnamefont
  {Androić}} \emph {et~al.} (\bibinfo {collaboration} {Qweak}),\ }\href
  {\doibase 10.1038/s41586-018-0096-0} {\bibfield  {journal} {\bibinfo
  {journal} {Nature}\ }\textbf {\bibinfo {volume} {557}},\ \bibinfo {pages}
  {207} (\bibinfo {year} {2018})},\ \Eprint {http://arxiv.org/abs/1905.08283}
  {arXiv:1905.08283 [nucl-ex]} \BibitemShut {NoStop}%
\bibitem [{\citenamefont {Safronova}\ \emph {et~al.}(2018)\citenamefont
  {Safronova}, \citenamefont {Budker}, \citenamefont {DeMille}, \citenamefont
  {Kimball}, \citenamefont {Derevianko},\ and\ \citenamefont
  {Clark}}]{Safronova:2017xyt}%
  \BibitemOpen
  \bibfield  {author} {\bibinfo {author} {\bibfnamefont {M.~S.}\ \bibnamefont
  {Safronova}}, \bibinfo {author} {\bibfnamefont {D.}~\bibnamefont {Budker}},
  \bibinfo {author} {\bibfnamefont {D.}~\bibnamefont {DeMille}}, \bibinfo
  {author} {\bibfnamefont {D.~F.~J.}\ \bibnamefont {Kimball}}, \bibinfo
  {author} {\bibfnamefont {A.}~\bibnamefont {Derevianko}}, \ and\ \bibinfo
  {author} {\bibfnamefont {C.~W.}\ \bibnamefont {Clark}},\ }\href {\doibase
  10.1103/RevModPhys.90.025008} {\bibfield  {journal} {\bibinfo  {journal}
  {Rev. Mod. Phys.}\ }\textbf {\bibinfo {volume} {90}},\ \bibinfo {pages}
  {025008} (\bibinfo {year} {2018})},\ \Eprint
  {http://arxiv.org/abs/1710.01833} {arXiv:1710.01833 [physics.atom-ph]}
  \BibitemShut {NoStop}%
\bibitem [{\citenamefont {Davoudiasl}\ \emph
  {et~al.}(2012{\natexlab{a}})\citenamefont {Davoudiasl}, \citenamefont {Lee},\
  and\ \citenamefont {Marciano}}]{Davoudiasl:2012ag}%
  \BibitemOpen
  \bibfield  {author} {\bibinfo {author} {\bibfnamefont {H.}~\bibnamefont
  {Davoudiasl}}, \bibinfo {author} {\bibfnamefont {H.-S.}\ \bibnamefont {Lee}},
  \ and\ \bibinfo {author} {\bibfnamefont {W.~J.}\ \bibnamefont {Marciano}},\
  }\href {\doibase 10.1103/PhysRevD.85.115019} {\bibfield  {journal} {\bibinfo
  {journal} {Phys. Rev.}\ }\textbf {\bibinfo {volume} {D85}},\ \bibinfo {pages}
  {115019} (\bibinfo {year} {2012}{\natexlab{a}})},\ \Eprint
  {http://arxiv.org/abs/1203.2947} {arXiv:1203.2947 [hep-ph]} \BibitemShut
  {NoStop}%
\bibitem [{\citenamefont {Davoudiasl}\ \emph
  {et~al.}(2012{\natexlab{b}})\citenamefont {Davoudiasl}, \citenamefont {Lee},\
  and\ \citenamefont {Marciano}}]{Davoudiasl:2012qa}%
  \BibitemOpen
  \bibfield  {author} {\bibinfo {author} {\bibfnamefont {H.}~\bibnamefont
  {Davoudiasl}}, \bibinfo {author} {\bibfnamefont {H.-S.}\ \bibnamefont {Lee}},
  \ and\ \bibinfo {author} {\bibfnamefont {W.~J.}\ \bibnamefont {Marciano}},\
  }\href {\doibase 10.1103/PhysRevLett.109.031802} {\bibfield  {journal}
  {\bibinfo  {journal} {Phys. Rev. Lett.}\ }\textbf {\bibinfo {volume} {109}},\
  \bibinfo {pages} {031802} (\bibinfo {year} {2012}{\natexlab{b}})},\ \Eprint
  {http://arxiv.org/abs/1205.2709} {arXiv:1205.2709 [hep-ph]} \BibitemShut
  {NoStop}%
\bibitem [{\citenamefont {Davoudiasl}\ \emph {et~al.}(2015)\citenamefont
  {Davoudiasl}, \citenamefont {Lee},\ and\ \citenamefont
  {Marciano}}]{Davoudiasl:2015bua}%
  \BibitemOpen
  \bibfield  {author} {\bibinfo {author} {\bibfnamefont {H.}~\bibnamefont
  {Davoudiasl}}, \bibinfo {author} {\bibfnamefont {H.-S.}\ \bibnamefont {Lee}},
  \ and\ \bibinfo {author} {\bibfnamefont {W.~J.}\ \bibnamefont {Marciano}},\
  }\href {\doibase 10.1103/PhysRevD.92.055005} {\bibfield  {journal} {\bibinfo
  {journal} {Phys. Rev.}\ }\textbf {\bibinfo {volume} {D92}},\ \bibinfo {pages}
  {055005} (\bibinfo {year} {2015})},\ \Eprint
  {http://arxiv.org/abs/1507.00352} {arXiv:1507.00352 [hep-ph]} \BibitemShut
  {NoStop}%
\bibitem [{\citenamefont {Agostini}\ \emph {et~al.}(2017)\citenamefont
  {Agostini} \emph {et~al.}}]{Borexino:2017fbd}%
  \BibitemOpen
  \bibfield  {author} {\bibinfo {author} {\bibfnamefont {M.}~\bibnamefont
  {Agostini}} \emph {et~al.} (\bibinfo {collaboration} {Borexino}),\ }\href
  {\doibase 10.1103/PhysRevD.96.091103} {\bibfield  {journal} {\bibinfo
  {journal} {Phys. Rev.}\ }\textbf {\bibinfo {volume} {D96}},\ \bibinfo {pages}
  {091103} (\bibinfo {year} {2017})},\ \Eprint
  {http://arxiv.org/abs/1707.09355} {arXiv:1707.09355 [hep-ex]} \BibitemShut
  {NoStop}%
\bibitem [{\citenamefont {Beda}\ \emph {et~al.}(2013)\citenamefont {Beda},
  \citenamefont {Brudanin}, \citenamefont {Egorov}, \citenamefont {Medvedev},
  \citenamefont {Pogosov}, \citenamefont {Shevchik}, \citenamefont
  {Shirchenko}, \citenamefont {Starostin},\ and\ \citenamefont
  {Zhitnikov}}]{Beda:2013mta}%
  \BibitemOpen
  \bibfield  {author} {\bibinfo {author} {\bibfnamefont {A.~G.}\ \bibnamefont
  {Beda}}, \bibinfo {author} {\bibfnamefont {V.~B.}\ \bibnamefont {Brudanin}},
  \bibinfo {author} {\bibfnamefont {V.~G.}\ \bibnamefont {Egorov}}, \bibinfo
  {author} {\bibfnamefont {D.~V.}\ \bibnamefont {Medvedev}}, \bibinfo {author}
  {\bibfnamefont {V.~S.}\ \bibnamefont {Pogosov}}, \bibinfo {author}
  {\bibfnamefont {E.~A.}\ \bibnamefont {Shevchik}}, \bibinfo {author}
  {\bibfnamefont {M.~V.}\ \bibnamefont {Shirchenko}}, \bibinfo {author}
  {\bibfnamefont {A.~S.}\ \bibnamefont {Starostin}}, \ and\ \bibinfo {author}
  {\bibfnamefont {I.~V.}\ \bibnamefont {Zhitnikov}},\ }\href {\doibase
  10.1134/S1547477113020027} {\bibfield  {journal} {\bibinfo  {journal} {Phys.
  Part. Nucl. Lett.}\ }\textbf {\bibinfo {volume} {10}},\ \bibinfo {pages}
  {139} (\bibinfo {year} {2013})}\BibitemShut {NoStop}%
\bibitem [{\citenamefont {Fujikawa}\ and\ \citenamefont
  {Shrock}(1980)}]{PhysRevLett.45.963}%
  \BibitemOpen
  \bibfield  {author} {\bibinfo {author} {\bibfnamefont {K.}~\bibnamefont
  {Fujikawa}}\ and\ \bibinfo {author} {\bibfnamefont {R.~E.}\ \bibnamefont
  {Shrock}},\ }\href {\doibase 10.1103/PhysRevLett.45.963} {\bibfield
  {journal} {\bibinfo  {journal} {Phys. Rev. Lett.}\ }\textbf {\bibinfo
  {volume} {45}},\ \bibinfo {pages} {963} (\bibinfo {year} {1980})}\BibitemShut
  {NoStop}%
\bibitem [{\citenamefont {Schechter}\ and\ \citenamefont
  {Valle}(1981)}]{PhysRevD.24.1883}%
  \BibitemOpen
  \bibfield  {author} {\bibinfo {author} {\bibfnamefont {J.}~\bibnamefont
  {Schechter}}\ and\ \bibinfo {author} {\bibfnamefont {J.~W.~F.}\ \bibnamefont
  {Valle}},\ }\href {\doibase 10.1103/PhysRevD.24.1883} {\bibfield  {journal}
  {\bibinfo  {journal} {Phys. Rev. D}\ }\textbf {\bibinfo {volume} {24}},\
  \bibinfo {pages} {1883} (\bibinfo {year} {1981})}\BibitemShut {NoStop}%
\bibitem [{\citenamefont {Kayser}(1982)}]{Kayser:1982br}%
  \BibitemOpen
  \bibfield  {author} {\bibinfo {author} {\bibfnamefont {B.}~\bibnamefont
  {Kayser}},\ }\href {\doibase 10.1103/PhysRevD.26.1662} {\bibfield  {journal}
  {\bibinfo  {journal} {Phys. Rev.}\ }\textbf {\bibinfo {volume} {D26}},\
  \bibinfo {pages} {1662} (\bibinfo {year} {1982})}\BibitemShut {NoStop}%
\bibitem [{\citenamefont {Nieves}(1982)}]{PhysRevD.26.3152}%
  \BibitemOpen
  \bibfield  {author} {\bibinfo {author} {\bibfnamefont {J.~F.}\ \bibnamefont
  {Nieves}},\ }\href {\doibase 10.1103/PhysRevD.26.3152} {\bibfield  {journal}
  {\bibinfo  {journal} {Phys. Rev. D}\ }\textbf {\bibinfo {volume} {26}},\
  \bibinfo {pages} {3152} (\bibinfo {year} {1982})}\BibitemShut {NoStop}%
\bibitem [{\citenamefont {Pal}\ and\ \citenamefont
  {Wolfenstein}(1982)}]{PhysRevD.25.766}%
  \BibitemOpen
  \bibfield  {author} {\bibinfo {author} {\bibfnamefont {P.~B.}\ \bibnamefont
  {Pal}}\ and\ \bibinfo {author} {\bibfnamefont {L.}~\bibnamefont
  {Wolfenstein}},\ }\href {\doibase 10.1103/PhysRevD.25.766} {\bibfield
  {journal} {\bibinfo  {journal} {Phys. Rev. D}\ }\textbf {\bibinfo {volume}
  {25}},\ \bibinfo {pages} {766} (\bibinfo {year} {1982})}\BibitemShut
  {NoStop}%
\bibitem [{\citenamefont {Shrock}(1982)}]{Shrock:1982sc}%
  \BibitemOpen
  \bibfield  {author} {\bibinfo {author} {\bibfnamefont {R.~E.}\ \bibnamefont
  {Shrock}},\ }\href {\doibase 10.1016/0550-3213(82)90273-5} {\bibfield
  {journal} {\bibinfo  {journal} {Nucl. Phys.}\ }\textbf {\bibinfo {volume}
  {B206}},\ \bibinfo {pages} {359} (\bibinfo {year} {1982})}\BibitemShut
  {NoStop}%
\bibitem [{\citenamefont {Giunti}\ and\ \citenamefont
  {Studenikin}(2015)}]{Giunti:2014ixa}%
  \BibitemOpen
  \bibfield  {author} {\bibinfo {author} {\bibfnamefont {C.}~\bibnamefont
  {Giunti}}\ and\ \bibinfo {author} {\bibfnamefont {A.}~\bibnamefont
  {Studenikin}},\ }\href {\doibase 10.1103/RevModPhys.87.531} {\bibfield
  {journal} {\bibinfo  {journal} {Rev. Mod. Phys.}\ }\textbf {\bibinfo {volume}
  {87}},\ \bibinfo {pages} {531} (\bibinfo {year} {2015})},\ \Eprint
  {http://arxiv.org/abs/1403.6344} {arXiv:1403.6344 [hep-ph]} \BibitemShut
  {NoStop}%
\bibitem [{\citenamefont {Beda}\ \emph {et~al.}(2012)\citenamefont {Beda},
  \citenamefont {Brudanin}, \citenamefont {Egorov}, \citenamefont {Medvedev},
  \citenamefont {Pogosov}, \citenamefont {Shirchenko},\ and\ \citenamefont
  {Starostin}}]{Beda:2012zz}%
  \BibitemOpen
  \bibfield  {author} {\bibinfo {author} {\bibfnamefont {A.~G.}\ \bibnamefont
  {Beda}}, \bibinfo {author} {\bibfnamefont {V.~B.}\ \bibnamefont {Brudanin}},
  \bibinfo {author} {\bibfnamefont {V.~G.}\ \bibnamefont {Egorov}}, \bibinfo
  {author} {\bibfnamefont {D.~V.}\ \bibnamefont {Medvedev}}, \bibinfo {author}
  {\bibfnamefont {V.~S.}\ \bibnamefont {Pogosov}}, \bibinfo {author}
  {\bibfnamefont {M.~V.}\ \bibnamefont {Shirchenko}}, \ and\ \bibinfo {author}
  {\bibfnamefont {A.~S.}\ \bibnamefont {Starostin}},\ }\href {\doibase
  10.1155/2012/350150} {\bibfield  {journal} {\bibinfo  {journal} {Adv. High
  Energy Phys.}\ }\textbf {\bibinfo {volume} {2012}},\ \bibinfo {pages}
  {350150} (\bibinfo {year} {2012})}\BibitemShut {NoStop}%
\end{thebibliography}%

\end{document}